\documentclass[twocolumn,twocolappendix]{aastex63}

\usepackage{amsmath}
\usepackage{multirow}
\hypersetup{linkcolor=blue,citecolor=blue,filecolor=cyan,urlcolor=blue}
\graphicspath{{./}{figures/}}
\def \lamedd {$\lambda_{\mbox{\scriptsize Edd}}$}

\def \nh {$N_{\mbox{\scriptsize H}}$}
\def \cmmt {$\mbox{cm}^{-2}$}

\def \lognhtorcmmt {$\log\,N_{\mbox{\scriptsize H,tor}}/\mbox{cm}^{-2}$}
\def \nhtor {$N_{\mbox{\scriptsize H,tor}}$}

\def \lognhloscmmt {$\log\,N_{\mbox{\scriptsize H,los}}/\mbox{cm}^{-2}$}
\def \nhequ {$N_{\mbox{\scriptsize H,equ}}$}

\def \lognhequcmmt {$\log\,N_{\mbox{\scriptsize H,equ}}/\mbox{cm}^{-2}$}
\def \nhlos {$N_{\mbox{\scriptsize H,los}}$}

\def \rpex {$R_{\mbox{\scriptsize pex}}$}
\def \cftor {$C_{\mbox{\scriptsize tor}}$}
\def \dctor {$D_{\mbox{\scriptsize tor}}$}
\def \logdctor {$\log\,D_{\mbox{\scriptsize tor}}$}

\def \thetainc {$\theta_{\mbox{\scriptsize inc}}$}

\def \ecut {$E_{\mbox{\scriptsize cut}}$}
\def \pnull {$p_{\mbox{\scriptsize null}}$}
\def \fsca {$f_{\mbox{\scriptsize s}}$}
\def \feka {{Fe\,K$\alpha$}}

\def \fluxunit {erg\,s$^{-1}$\,cm$^{-2}$}

\def \swift {{Swift}}
\def \bepposax {{BeppoSAX}}
\def \integral {{INTEGRAL}}
\def \suzaku {{Suzaku}}
\def \nustar {{NuSTAR}}

\def \swiftbat {{Swift}/BAT}
\def \chandra {{Chandra}}
\def \rxte {{RXTE}}

\def \xmmnewton {{XMM-Newton}}

\def \borus {\texttt{BORUS}}
\def \borustwo {\texttt{borus02}}
\def \pexrav {\texttt{pexrav}}

\def \xspec {\texttt{Xspec}}

\shorttitle{Multi-epoch Broadband X-ray Spectroscopy of NGC\,1052}
\shortauthors{Balokovi\'{c} et al. (2021)}

\begin{document}

\title{Properties of the Obscuring Torus in NGC\,1052\\from Multi-epoch Broadband X-ray Spectroscopy}

\correspondingauthor{M.~Balokovi\'{c}}
\email{mislav.balokovic@yale.edu}

\author[0000-0003-0476-6647]{M.~Balokovi\'{c}}
\affiliation{Yale Center for Astronomy \& Astrophysics, 52 Hillhouse Avenue, New Haven, CT 06511, USA}
\affiliation{Department of Physics, Yale University, P.O. Box 2018120, New Haven, CT 06520, USA}
\affiliation{Black Hole Initiative at Harvard University, 20 Garden Street, Cambridge, MA 02138, USA}
\affiliation{Center for Astrophysics $\vert$ Harvard \& Smithsonian, 60 Garden Street, Cambridge, MA 02138, USA}
\author{S.~E.~Cabral}
\affiliation{Department of Physics, University of Massachusetts Boston, 100 Morrissey Blvd, Boston, MA 02125, USA}
\affiliation{Center for Astrophysics $\vert$ Harvard \& Smithsonian, 60 Garden Street, Cambridge, MA 02138, USA}
\affiliation{Black Hole Initiative at Harvard University, 20 Garden Street, Cambridge, MA 02138, USA}
\author[0000-0003-2663-1954]{L.~Brenneman}
\affiliation{Center for Astrophysics $\vert$ Harvard \& Smithsonian, 60 Garden Street, Cambridge, MA 02138, USA}
\author[0000-0002-0745-9792]{C.~M.~Urry}
\affiliation{Yale Center for Astronomy \& Astrophysics, 52 Hillhouse Avenue, New Haven, CT 06511, USA}
\affiliation{Department of Physics, Yale University, P.O. Box 2018120, New Haven, CT 06520, USA}

\begin{abstract} 
Obscuration of the innermost parts of active galactic nuclei (AGN) is observed in the majority of the population both in the nearby universe and at high redshift. However, the nature of the structures causing obscuration, especially in low-luminosity AGN, is poorly understood at present. We present a novel approach to multi-epoch broadband X-ray spectroscopy, anchored in the long-term average spectrum in the hard X-ray band, applied to the nearby, X-ray bright AGN in the galaxy NGC\,1052. From spectral features due to X-ray reprocessing in the circumnuclear material, based on a simple, uniform-density torus X-ray reprocessing model, we find a covering factor of 80--100\,\% and a globally averaged column density in the range $(1-2)\times10^{23}$\,\cmmt. This closely matches the independently measured variable line-of-sight column density range, leading to a straightforward and self-consistent picture of the obscuring torus in NGC\,1052, similar to several other AGN in recent literature. Comparing this X-ray-constrained torus model with measurements of spatially resolved sub-parsec absorption from radio observations, we find that it may be possible to account for both X-ray and radio data with a torus model featuring a steep density gradient along the axis of the relativistic jets. This provides a valuable direction for the development of improved physical models for the circumnuclear environment in NGC\,1052 and potentially in a wider class of AGN.
\end{abstract} 


\section{Introduction} 
\label{sec:intro} 

In the framework of the simple unified model of active galactic nuclei (AGN), an anisotropic obscuring structure is needed to explain obscuration in the X-ray band and the dichotomy of optical types \citep{antonucci-1993,urry+padovani-1995}, among other phenomenological features. This structure is traditionally called the ``torus'' even though in reality it is likely more complicated than its geometrical namesake, possibly a combination of several dynamical structures \citep{netzer-2015,hoenig-2019}. The gas and dust in the torus absorb and reprocess radiation from the innermost regions around the supermassive black hole (SMBH), creating observable spectral signatures accessible to current instruments in the infrared \citep[e.g.,][]{alonsoHerrero+2011,garciaBernete+2019} and X-ray bands \citep[e.g.,][]{murphy+yaqoob-2009,liu+li-2014}.

As seen from the SMBH, the torus is thought to cover a significant fraction of outward lines of sight with column density (\nh) above $10^{24}$\,\cmmt\ \citep{ramosAlmeida+ricci-2017}, where the material becomes Compton-thick (CT) as the optical depth to Compton scattering exceeds unity. Recent results from broadband X-ray spectroscopy of bright, nearby AGN suggest that signatures of X-ray reprocessing in the torus do not require the presence of CT material outside of our line of sight to the nucleus, both in obscured \citep[e.g.,][]{yaqoob+2015-mrk3,zhao+2020} and unobscured cases \citep[e.g.,][]{ursini+2015-ngc7213,younes+2019,diaz+2020-ngc3718}. However, the simplicity of the currently available spectral models for X-ray reprocessing in the torus leaves open the possibility that the torus contains clumps of CT material, while the \nh\ averaged over the torus is significantly lower \citep{balokovic-2017,zhao+2021}. Observations of variability in the line-of-sight column density, \nhlos\ \citep[e.g.,][]{risaliti+2002,guainazzi+2016-mrk3,zaino+2020-ngc1068}, are a powerful probe of torus clumpiness \citep{markowitz+2014,buchner+2019,laha+2020}.

In this paper, we focus on the properties of the obscuring torus in the bright, nearby AGN at the center of the galaxy NGC\,1052. It exhibits a type~2 optical spectrum with broad lines observed in polarized light \citep{barth+1999-ngc1052}. Studying its emission across the electromagnetic spectrum, \citet{fernandezOntiveros+2019-ngc1052} estimated that its bolometric luminosity ($L_{\rm bol}$) and Eddington ratio ($\lambda_{\rm Edd}=L_{\rm bol}/L_{\rm Edd}$) are around $7\times10^{42}$\,\,erg\,s$^{-1}$ and $4\times10^{-4}$, respectively. The relatively low accretion rate classifies NGC\,1052 as a low-luminosity AGN (LL\,AGN; \citealt{ho-2008}), although it has previously been found to share some characteristics of typically more luminous Seyfert galaxies. Recent work suggests that the X-ray spectrum may be one of those characteristics \citep{brenneman+2009-ngc1052,osorioClavijo+2020-ngc1052,cabral-2020}. In contrast, earlier studies \citep{weaver+1999-ngc1052,guainazzi+2000-ngc1052,kadler+2004a-ngc1052-mwl} argued for a significantly harder intrinsic continuum more similar to that of jet-dominated sources \citep[e.g.,][]{sambruna+2006,gianni+2011}.

NGC\,1052 features a pair of relativistic jets which have been characterized in detail using Very Long Baseline Interferometry (VLBI) up to high frequencies in the radio band \citep[see][and references therein]{baczko+2019-ngc1052,nakahara+2020-ngc1052}. These observations provide a useful constraint on the inclination of the system, likely in the range 60--85$^{\circ}$ \citep{kadler+2004b-ngc1052-vlbi,baczko+2016-ngc1052}. VLBI observations also established the existence of free-free absorption in the medium surrounding the innermost part of the twin jets at sub-pc scales \citep{kameno+2001-ngc1052,sawadaSatoh+2008-ngc1052,baczko+2016-ngc1052}, which is typically ascribed to a partially ionized torus. Very high opacity ($\tau\simeq1000$) has been observed within about 2\,mas ($\simeq0.2$\,pc) from the estimated position of the SMBH, obscuring the receding jet \citep{sawadaSatoh+2008-ngc1052}. Multiple molecular species have been observed as masers or absorbers at comparable angular scale \citep{claussen+1998-ngc1052,vermeulen+2003-ngc1052,impellizzeri+2008-ngc1052}, leading to estimates of \nh\ in the CT regime in at least some parts of the obscuring torus \citep{sawadaSatoh+2016-ngc1052,sawadaSatoh+2019-ngc1052}.

The goal of this study is to investigate the constraints on the basic parameters of the obscuring torus in NGC\,1052 using X-ray data. To this end, we take a novel approach to multi-epoch broadband X-ray spectroscopy anchored in the long-term average spectrum in the hard X-ray band ($>\!10$\,keV), for which we make use of all currently available hard X-ray data. The hard X-ray band is critical for determining the properties of the intrinsic continuum, which is substantially affected by absorption at $<\!10$\,keV. It is also essential for characterizing the Compton hump, as the key spectral signature (in addition to the \feka\ line) of circumnuclear reprocessing associated with the torus. We first establish a baseline spectral model using the highest-quality single-epoch data in \S\,\ref{sec:fitting-singleepoch}, then expand the analysis to multiple epochs in \S\,\ref{sec:fitting-multiepoch}. Folding in the analysis of obscuration variability (\S\,\ref{sec:fitting-lc}), we construct a self-consistent geometrical model for the obscuring torus in NGC\,1052, which we further discuss and compare to sub-pc radio opacity measurements in \S\,\ref{sec:discussion}.

\section{Observations and Data} 
\label{sec:obs} 

In this paper, we specifically focus on maximizing the hard X-ray coverage in order to obtain the best possible constraints on the part of the X-ray spectrum that is essentially unaffected by the effects of photoelectric absorption. Much of the existing soft X-ray data has previously been analyzed and presented in the literature. In particular, soft X-ray diffuse emission in NGC\,1052 has been studied using \chandra\ data by \citet{kadler+2004a-ngc1052-mwl} and most recently by \citet{osorioClavijo+2020-ngc1052} and \citet{falocco+2020-ngc1052}. These works, along with some studies of larger samples that include NGC\,1052 \citep{hernandezGarcia+2013-liners,connolly+2016}, also examined variability in the soft X-ray band. Individual hard X-ray spectra have previously been analyzed by \citet{guainazzi+2000-ngc1052}, \citet{brenneman+2009-ngc1052}, and \citet{rivers+2013}, while \citet{osorioClavijo+2020-ngc1052} and \citet{cabral-2020} presented multi-epoch studies based on some of the available hard X-ray data combined with different sets of soft X-ray observations. Individual pointed observations we consider in this work are listed in Table~\ref{tab:obslog}.

\begin{deluxetable*}{cccccccc}
\tablecaption{ Individual X-ray observations of NGC\,1052 considered in this work \label{tab:obslog}}
\tabletypesize{\small}
\tablehead{
  \colhead{Epoch} &
  \colhead{Observatory} &
  \colhead{Observation ID} &
  \colhead{Start Date} &
  \colhead{Instrument} &
  \colhead{Exposure (ks)} &
  \colhead{Band (keV)} &
  \colhead{Count Rate\,\tablenotemark{a} (s$^{-1}$)}
}
\startdata
\multirow{5}{*}{1}
 & \multirow{2}{*}{\nustar} & \multirow{2}{*}{2017-01-01} & \multirow{2}{*}{60201056002} & FPMA & 55.4 & \multirow{2}{*}{3--79} & $0.214\pm0.002$ \\
 & & & & FPMB & 55.9 & & $0.200\pm0.002$ \\
 & \multirow{3}{*}{\xmmnewton} & \multirow{3}{*}{2017-01-01} & \multirow{3}{*}{0790980101} & EPIC/PN & 43.4 & 0.5--10 & $0.502\pm0.003$ \\
 & & & & EPIC/MOS1\,\tablenotemark{b} & 55.4 & \multirow{2}{*}{0.5--10} & \multirow{2}{*}{$0.148\pm0.001$} \\
 & & & & EPIC/MOS2\,\tablenotemark{b} & 58.2 & \\
\hline
\multirow{2}{*}{2}
 & \multirow{2}{*}{\nustar} & \multirow{2}{*}{2013-02-14} & \multirow{2}{*}{60061027002} & FPMA & 15.6 & \multirow{2}{*}{3--79} & $0.171\pm0.003$ \\
 & & & & FPMB & 15.6 & & $0.171\pm0.003$ \\
\hline
\multirow{4}{*}{3}
 & \multirow{4}{*}{\suzaku} & \multirow{4}{*}{2007-02-16} & \multirow{4}{*}{702058010} & HXD/PIN & 78.1 & 12--55 & $0.039\pm0.003$ \\
 & & & & XIS1 & 100.1 & 0.7--7 & $0.107\pm0.001$ \\
 & & & & XIS0\,\tablenotemark{b} & 100.1 & \multirow{2}{*}{0.7--7} & \multirow{2}{*}{$0.119\pm0.001$} \\
 & & & & XIS3\,\tablenotemark{b} & 100.1 & & \\
\hline
\multirow{4}{*}{4}
 & \multirow{4}{*}{\bepposax} & \multirow{4}{*}{2000-01-11} & \multirow{4}{*}{5082800} & PDS & 30.0 & 15--180 & $0.20\pm0.04$ \\
 & & & & MECS2\,\tablenotemark{b} & 31.8 & \multirow{2}{*}{2--9} & \multirow{2}{*}{$0.0312\pm0.0007$} \\
 & & & & MECS3\,\tablenotemark{b} & 31.8 & & \\
 & & & & LECS & 25.7 & 2--9 & $0.0102\pm0.0007$ \\
\hline
\enddata
\tablenotetext{a}{Background-subtracted source count rate in the given energy band, without PSF corrections.}
\tablenotetext{b}{Data for two similar instruments coadded for the spectral analysis.}
\end{deluxetable*}

\subsection{Long-term Light Curves and Average Spectra} 
\label{sec:obs-longterm} 

\begin{figure}
\begin{center}
\includegraphics[width=\columnwidth]{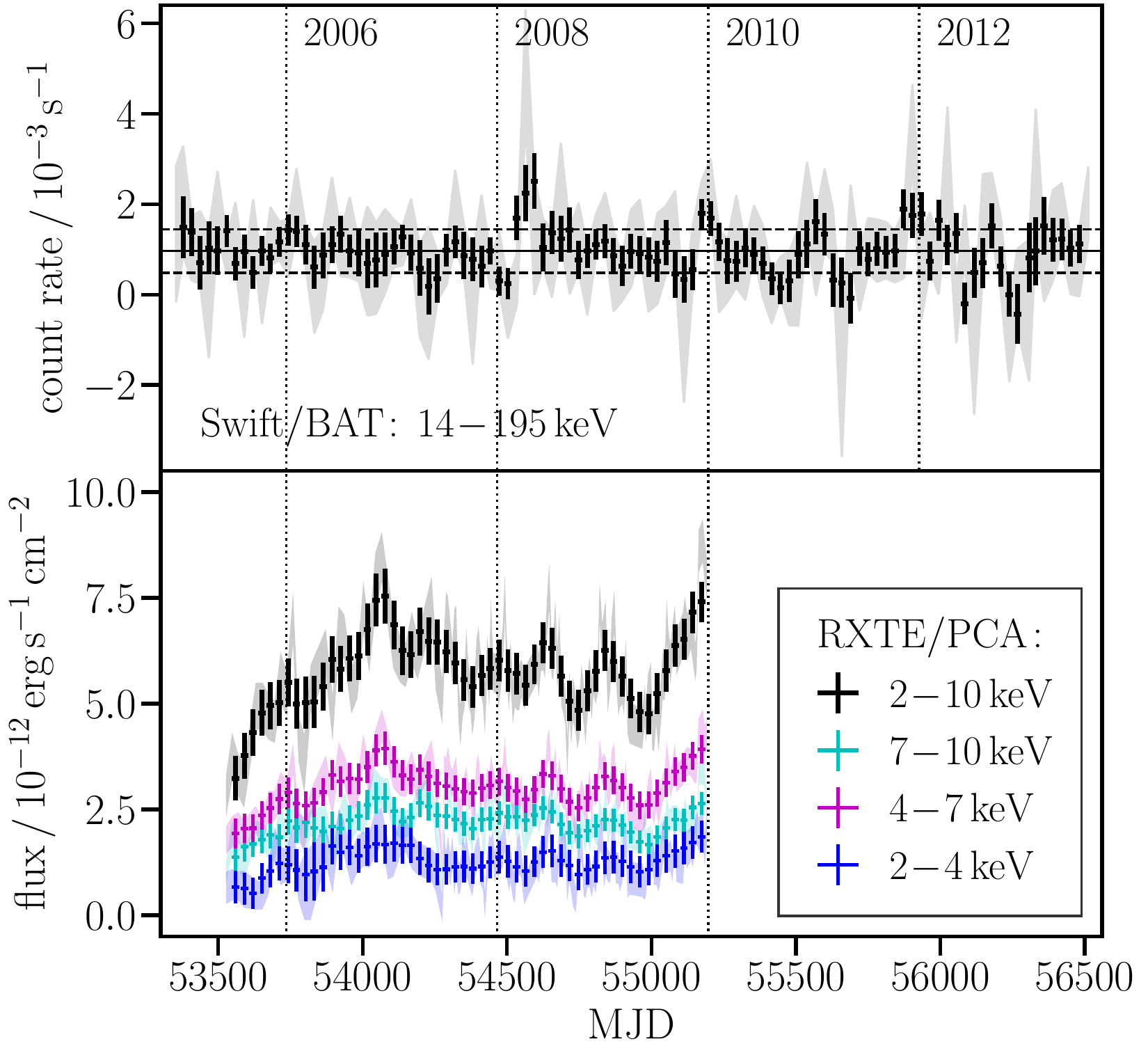}
\caption{ Light curves in the hard \swiftbat\ band (upper panel) and in several soft \rxte/PCA sub-bands (lower panel). In all cases, we show the original light curves as lightly shaded regions, while the points with error bars show three-month averages plotted with a one-month cadence. In the upper panel, the solid black horizontal line shows the average count rate with the dashed lines indicating a 50\,\% departure from the average. Vertical dotted lines mark starts of even years 2006--2012. \label{fig:lcs}}
\end{center}
\end{figure}

NGC\,1052 is a relatively bright hard X-ray source, and has been detected by the Burst Alert Telescope (BAT; \citealt{barthelmy+2005-bat}) on board the {\em Neil Gerhels Swift Observatory} (\swift\ hereafter; \citealt{gehrels+2004}) in the 14--195\,keV band. It was first included in the \swiftbat\ all-sky survey catalog in its 22-month edition \citep{tueller+2010-bat22}. In the latest 105-month edition \citep{oh+2018-bat105}, NGC\,1052 is detected with a signal-to-noise ratio of 15, and its listed 14--195\,keV flux is $(3.1\pm0.3)\times10^{-11}$\,\fluxunit. In this work we make use of the light curve and spectrum constructed from data accumulated between December 2004 and September 2013 available from the 105-month online catalog.\footnote{\url{https://swift.gsfc.nasa.gov/results/bs105mon}} We show the light curve in the top panel of Figure~\ref{fig:lcs}.

The {\em Rossi X-ray Timing Explorer} (\rxte; \citealt{bradt+1993-rxte}) observed NGC\,1052 on about 150 occasions between June 2005 and December 2009. These observations have been analyzed as a part of the {\em RXTE AGN Timing \& Spectral Database} \citep{rivers+2013} with data products available online.\footnote{\url{http://cass.ucsd.edu/~rxteagn}} In total, NGC\,1052 has been observed for approximately 400\,ks with the PCA instrument \citep{jahoda+2006-rxtepca}, covering the 2--60\,keV band. In our analysis, we make use of the average spectrum as well as the light curves in 2--4\,keV, 4--7\,keV, and 7--10\,keV bands. These light curves are provided in \fluxunit, calculated from a spectral model fitted to the 3.3--10\,keV band (as the source is not significantly detected at lower energies) and are shown in the lower panel of Figure~\ref{fig:lcs}.

We also use the hard X-ray spectrum (20--150\,keV) constructed from data collected by \integral\ \citep{winkler+2003-integral}. We obtained the spectrum from the third revision of the online archive\footnote{\url{https://www.isdc.unige.ch/integral/heavens}} hosted by the {\em INTEGRAL Science Data Centre}. It is built from observations with the IBIS/ISGRI instrument \citep{lebrun+2003-isgri,ubertini+2003-ibis} between 2003 and 2013 following the data processing procedure is described by \citet{walter+2010-isdc}.

The light curves shown in Figure~\ref{fig:lcs} demonstrate that the source is approximately steady around the long-term average flux in the hard X-ray \swiftbat\ band, while at the same time \rxte/PCA sampled more significant variability below 10\,keV. To the original statistical uncertainties in the soft X-ray light curves we added systematic uncertainties due to possibly variable background estimated to be $2-4\times10^{-13}$\,\fluxunit\ (2--4\,keV), $1-2\times10^{-13}$\,\fluxunit\ (4--7\,keV), and $1\times10^{-13}$\,\fluxunit\ (7--10\,keV), as noted in the online database. In order to make the light curves more directly comparable and reduce uncertainties, we re-binned them on the same monthly time grid while averaging over 3-month periods.

\begin{figure}
\begin{center}
\includegraphics[width=\columnwidth]{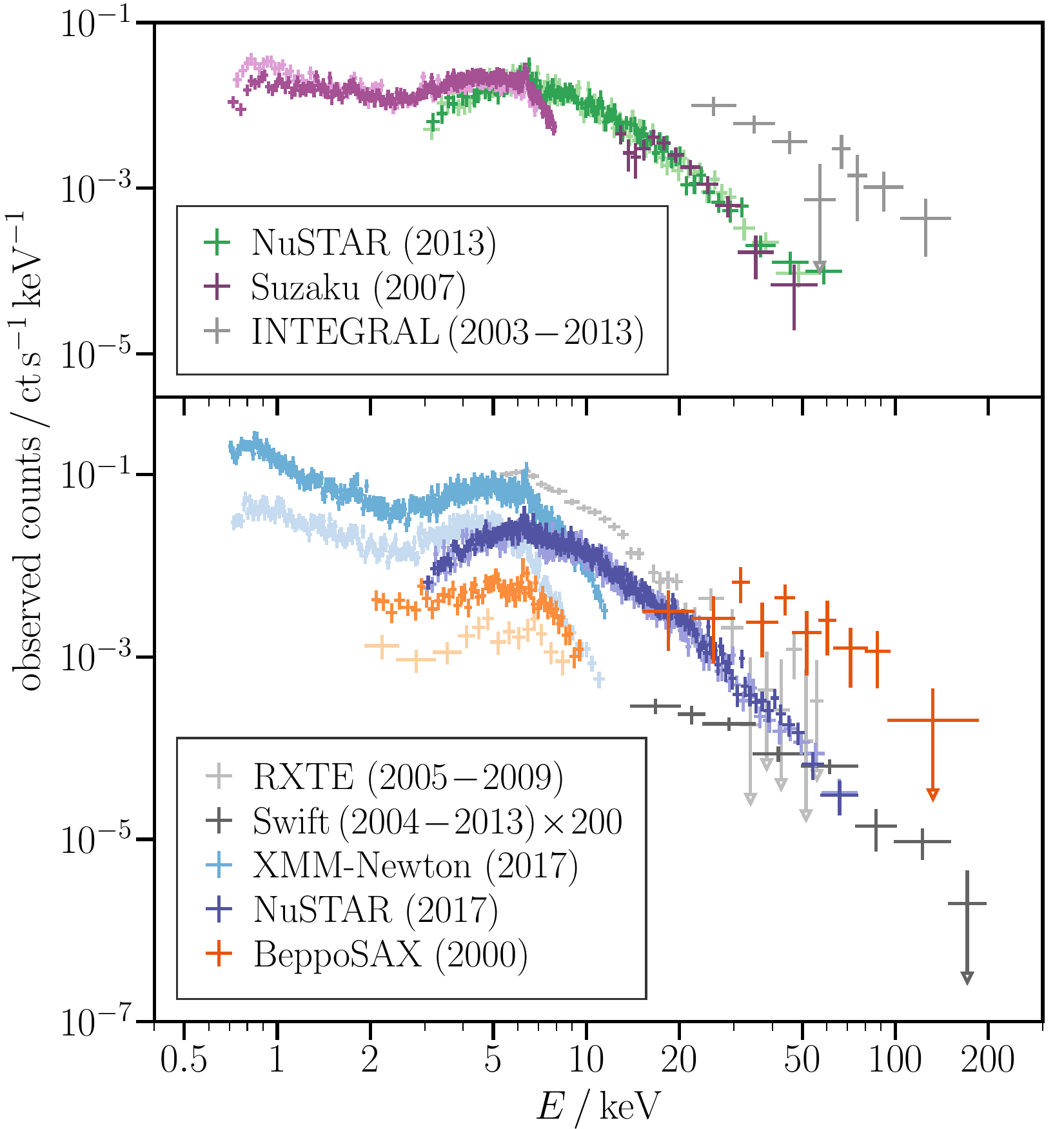}
\caption{ Raw counts spectra for all observations considered in this work over the energy ranges used in our multi-epoch spectral modeling. The scaling of the vertical axis is the same in both panels; they are separated only for clarity. The \swiftbat\ spectrum is shown multiplied by a factor of 200 because of an otherwise large downward offset from the other spectra. Downward-pointing arrows mark bins with error bars formally extending to zero. \label{fig:counts}}
\end{center}
\end{figure}

All three spectra are integrated over a number of years but still feature limited photon statistics in their highest-energy bins (see the grey spectra in Figure~\ref{fig:counts}). We use them without re-binning, effectively treating some of the highest-energy bins as upper limits in our spectral fitting. Including those bins does not have a significant effect on the key parameters of our broadband spectral model. Considered alone, the BAT, ISGRI, and PCA spectra agree in the overlapping 20--60\,keV band in terms of the effective photon index ($\Gamma_{\rm \scriptsize [20,60]}=1.8\pm0.3$, $1.8\pm0.6$, and $1\pm1$, respectively) and flux, which is within 20\,\% of the average $1.2\times10^{-11}$\,\fluxunit. In all three cases, we make use of the response files provided by the respective online databases.

\subsection{NuSTAR} 
\label{sec:obs-nustar} 

\nustar\ \citep{harrison+2013} observed NGC\,1052 on two occasions, 2013 February~14 (obsID 60061027002) and 2017 January~1 (obsID 60201056002). The data reduction and analysis followed the standard procedure outlined in the {\em NuSTAR Data Analysis Software Guide}\footnote{\url{https://heasarc.gsfc.nasa.gov/docs/nustar/analysis/nustar_swguide.pdf}} for each telescope (FPMA and FPMB) separately. We used the \nustar\ data analysis software package, NuSTARDAS version 1.8.0, provided within HEASOFT version 6.26, along with the \nustar\ calibration database version 20190430. After filtering the raw data using the \texttt{nupipeline} task (with \texttt{saa=optimized} and \texttt{tentacle=yes} options), the exposure times are 15.6\,ks and 55.7\,ks, respectively. The source does show fluctuations in count rate at about the 20\,\% level during the course of the longer observation, but we only make use of observation-averaged data products in this work. We extracted source spectra from circular regions centered on NGC\,1052 with a radius of 60\arcsec. Background spectra were extracted from source-free regions covering the same chip as the source for each observation, excluding circular regions within 90\arcsec\ from the source. The \texttt{nuproducts} task was used to generate source and background spectra along with the response files. We binned the spectra using the scripted procedure described in \citet{balokovic-2017}, which results in an approximately constant signal-to-noise ratio (SNR) per bin, with a minimum at SNR$>\!3$.

\subsection{XMM-Newton} 
\label{sec:obs-xmm} 

Our target was observed by \xmmnewton\ \citep{jansen+2001-xmm} multiple times, one of which was essentially simultaneous with the longer \nustar\ observation on 2017 January 1 (obsID 0790980101). We consider only this one epoch in our analysis, as the two observatories jointly provide a snapshot of the broadband X-ray spectrum of NGC\,1052 (0.5--79\,keV) with unparalleled sensitivity. We processed the data using the \xmmnewton\ {\em Science Analysis System} \citep{gabriel+2004-xmmsas} version 17.0, following the standard procedures outlined in the {\em XMM-Newton ABC Guide}.\footnote{\url{https://heasarc.gsfc.nasa.gov/docs/xmm/abc/}} For the EPIC instrument detector PN \citep{strueder+2001-epicpn} we selected only single and double-patterned events, while for the MOS detectors \citep{turner+2001-epicmos} we also included quadruple-patterned ones. We excluded intervals of relatively high background count rates that exceeded a factor of 2 above the non-flaring rates. After filtering the raw event files for PN, MOS1, and MOS2, the exposure times were 43.4\,ks, 55.4\,ks, and 58.2\,ks, respectively. The source data were extracted from circular regions with a 30\arcsec\ radius centered on NGC\,1052. Background regions were extracted from larger nearby regions within the same chip avoiding any chip gaps and faint point sources. For each detector, response files were generated using tasks \texttt{rmfgen} and \texttt{arfgen}, after which spectral and response files for MOS were combined using the \texttt{addascaspec} script. Finally, we binned the source spectra to have at least 50 counts per bin.

\subsection{Suzaku} 
\label{sec:obs-suzaku} 

\suzaku\ \citep{mitsuda+2007-suzaku} observed NGC\,1052 on 2007 February 16 (obsID 702058010). At that time, \suzaku\ had three operational XIS telescopes \citep{koyama+2007-suzakuxis} covering the soft X-ray band, and the target was significantly detected ($>\!10\,\sigma$) with the non-focusing hard X-ray instrument HXD/PIN \citep{takahashi+2007-suzakuhxd}. We processed the raw data using HEASOFT version 6.18 following standard procedures described in the {\em Suzaku ABC Guide}.\footnote{\url{https://heasarc.gsfc.nasa.gov/docs/suzaku/analysis/abc/}} After filtering, the effective exposure times were 100.1\,ks for each of the three XIS detectors (XIS0, XIS1, and XIS3) and 78.1\,ks for the HXD/PIN. The XIS source spectra were extracted from circular regions 3\arcmin\ in radius, combining {{$3\!\times\!3$}} and {{$5\!\times\!5$}} modes. Background spectra were extracted from large, source-free circular areas for each XIS detector away from any chip edges and calibration sources. Response files for each detector were generated using the tasks \texttt{xisrmfgen} and \texttt{xissimarfgen}. The spectra and response files from the two front-illuminated chips (XIS0 and XIS3) were then combined using \texttt{addascaspec}. We binned the spectra to a minimum of 100 counts per bin in order to roughly match the number of bins in the \xmmnewton\ spectra. We used the \texttt{hxdpinxbpi} script to generate response files and the background spectrum for PIN data, including both the instrumental and the cosmic X-ray background contributions. The PIN spectrum is background-dominated, so we grouped the data with a minimum of 3000 counts per bin.

\subsection{BeppoSAX} 
\label{sec:obs-bepposax} 

NGC\,1052 was observed with \bepposax\ \citep{boella+1997-bepposax} on 2000 January 11 (obsID 5082800). We obtained fully processed data for this observation from the HEASARC archive,\footnote{\url{https://heasarc.gsfc.nasa.gov/docs/archive.html}} produced following the standard procedures detailed in the \bepposax\ {\em ABC Guide}.\footnote{\url{https://heasarc.gsfc.nasa.gov/docs/sax/abc/saxabc/saxabc.html}} The data products include spectra from the soft X-ray telescopes LECS \citep{parmar+1997-bepposaxlecs} and MECS \citep{boella+1997-bepposaxmecs} extracted from circular regions with 3\arcmin\ radii around the source centroid. For the latter, data from two detector units (MECS2 and MECS3) operational at the time of the observation were combined into a single spectrum. The target was also significantly detected at hard X-ray energies using the non-focusing PDS instrument \cite{frontera+1997-bepposaxpds}. Total exposure times for the LECS, MECS (combined), and PDS instruments are 25.7\,ks, 63.7\,ks, and 30.0 ks, respectively. Appropriate response and background files were downloaded from the Italian Space Agency (ASI) online repository.\footnote{\url{ftp://ftp.asdc.asi.it/sax/cal}} We applied binning of at least 30 counts per bin for the soft X-ray instruments, and at least 500 counts per bin for the background-dominated PDS.

\section{Data Analysis} 
\label{sec:modeling} 

Data analysis presented here was performed using \xspec\ \citep{arnaud-1996} version 12.9.1m. All spectral models include an absorption component due to the Milky Way with the line-of-sight column density of $2.8\times10^{20}$\,\cmmt\ \citep{hi4pi-2016-nhgal} and $z=0.00504$ based on the heliocentric velocity measured by \citet{denicolo+2005}. We use $\chi^2$ statistics for fitting and evaluation of spectral models, choosing null hypothesis probability (\pnull) threshold of 5\,\% to accept or reject a particular model. In all fits we assume unity cross-normalization factor for one detector per epoch (see Table ~\ref{tab:obslog} for the definition of epochs), while others are optimized in the fitting procedure. A brief discussion of the cross-normalization strategy is given in the Appendix.

\subsection{Single-epoch Spectral Modeling} 
\label{sec:fitting-singleepoch} 

\begin{figure}
\begin{center}
\includegraphics[width=\columnwidth]{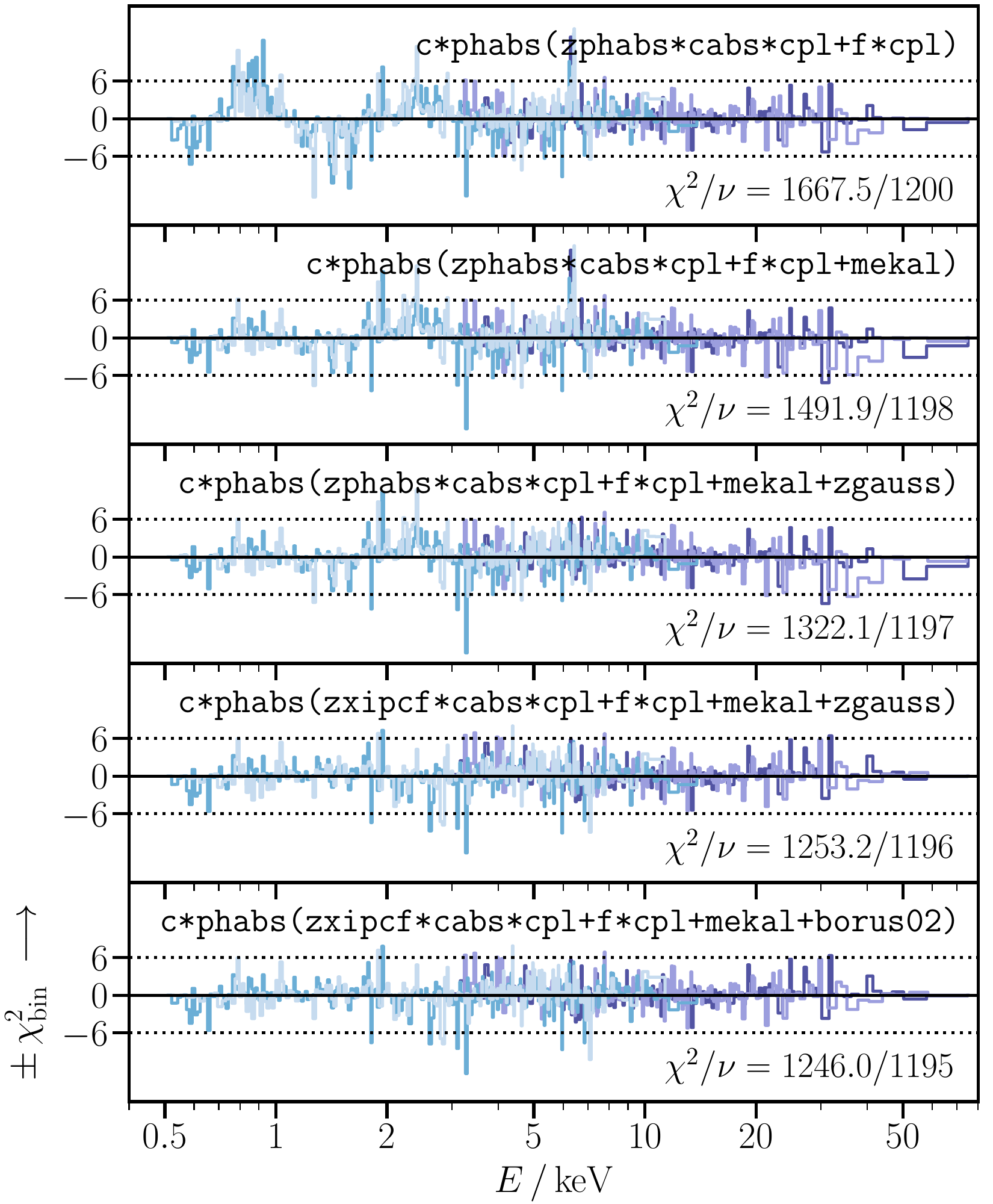}
\vspace{-0.5cm}
\caption{ Residuals in terms of $\chi^2$ contributions of individual energy bins in \xmmnewton\ and \nustar\ spectra from 2017 (epoch~1) for a series of models discussed in \S\,\ref{sec:fitting-singleepoch}. Shortened \xspec\ model expressions and total $\chi^2$ over the number of degrees of freedom ($\nu$) are given in each panel. For the former, \texttt{c} and \texttt{f} refer to the instrumental cross-normalization factor and the parameter \fsca, respectively, while \texttt{cpl} represents the cutoff power law model \texttt{cutoffpl} and the \borustwo\ component is included in the form of an additive table. \label{fig:sepchis}}
\end{center}
\end{figure}

\begin{figure}
\begin{center}
\includegraphics[width=\columnwidth]{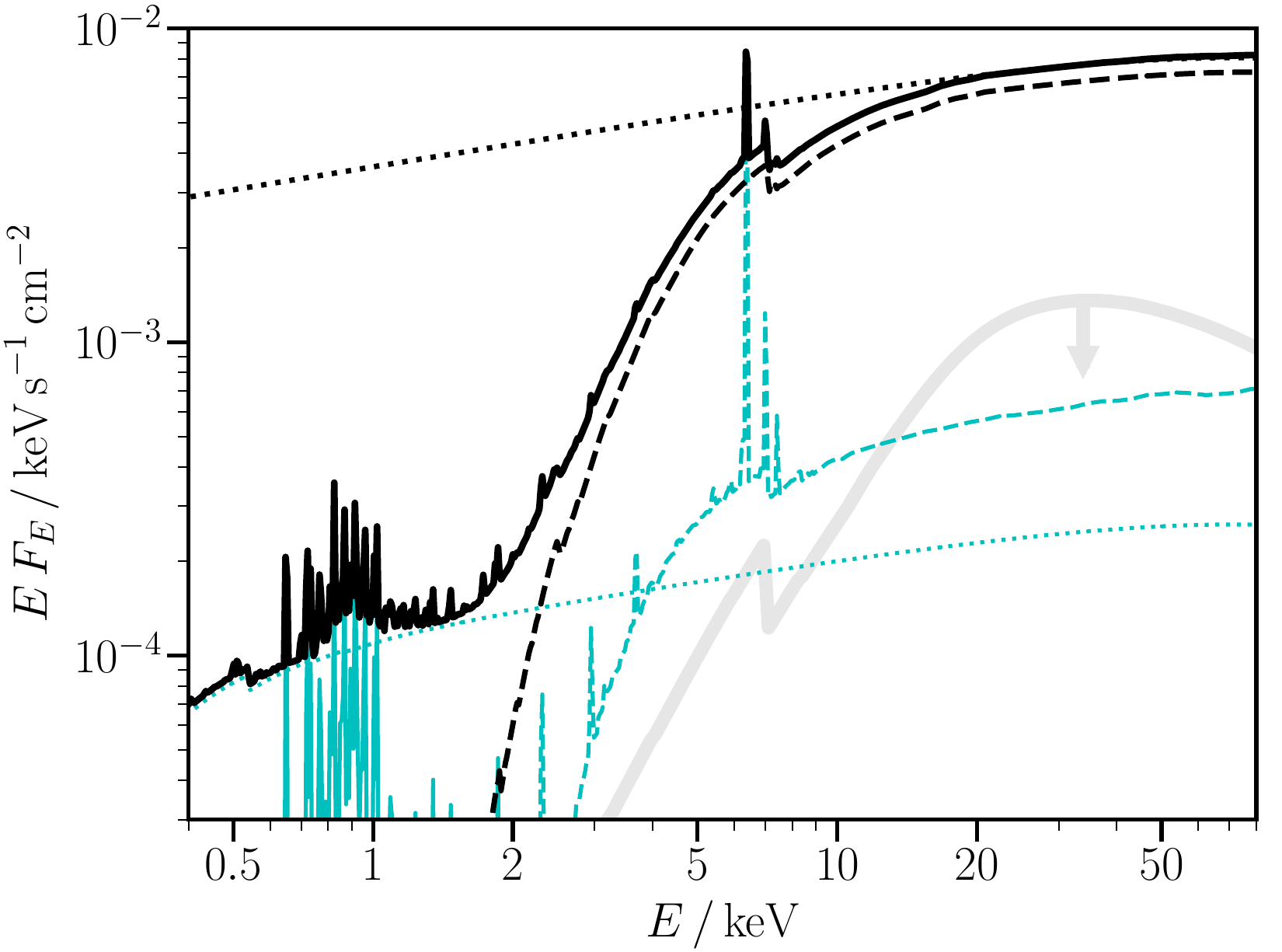}
\vspace{-0.6cm}
\caption{ Best-fit spectral model~S based on data from quasi-simultaneous \nustar\ and \xmmnewton\ observation in 2017. The solid black line shows the total. Dotted and dashed black lines show the intrinsic continuum and the component absorbed along the line of sight, respectively. The cyan lines show components not originating in the line of sight to the central source: \texttt{mekal} in solid, secondary power law in dotted, and \borustwo\ in dashed lines. For visual comparison, we also show the shape of the reprocessed continuum represented by \pexrav\ \citep{magdziarz+zdziarski-1995} with the thick, solid, light grey line, plotted here at the upper limit of its possible contribution according to the recent literature: $\vert\,$\rpex$\,\vert <0.3$ from \citet{balokovic-2017}, consistent with $\simeq0.1$ from \citet{osorioClavijo+2020-ngc1052} and $<0.01$ from \citet{brenneman+2009-ngc1052}. \label{fig:eemod}}
\end{center}
\end{figure}

As a first step in our multi-epoch analysis, we make use of the highest-quality broadband snapshot of the NGC\,1052 spectrum acquired in joint observation with \xmmnewton\ and \nustar\ in 2017. We start our analysis assuming a double power-law model with a high-energy cutoff and neutral line-of-sight absorption, following the basic steps from the most recent analyses of NGC\,1052 broadband X-ray spectra \citep{brenneman+2009-ngc1052,osorioClavijo+2020-ngc1052,cabral-2020}. We keep the coronal high-energy cutoff fixed at 290\,keV, as the representative median for the nearby obscured AGN population \citep{balokovic+2020}. Leaving out the Galactic absorption and the cross-normalization factor, the \xspec\ expression for this model is \texttt{zphabs}$\times$\texttt{cabs}$\times$\texttt{cutoffpl}+\fsca$\times$\texttt{cutoffpl}. The first term represents the intrinsic (primary) continuum with the AGN-related photoelectric absorption including line-of-sight Compton scattering. The secondary continuum component, observable only in the soft X-ray band is normalized relative to the primary continuum via the free parameter \fsca. Other free parameters are the normalization and the photon index ($\Gamma$) of the primary continuum, line-of-sight absorption column (\nhlos), and three cross-normalization factors (for FPMB, PN and coadded MOS spectra).

The first model does not fit the data well, with the lowest $\chi^2$ exceeding 1600 for 1200 degrees of freedom ($\nu$). Residuals in terms of $\chi^2$ contributions from each energy bin are shown in the top panel of Figure~\ref{fig:sepchis}. The low-energy excess peaking around 0.9\,keV may be due to optically thin plasma emission, while the narrow excess at 6.4\,keV reveals the presence of a narrow emission line. Both features have been identified in previous studies of NGC\,1052, with the former known to extend out to galactic scales \citep{kadler+2004a-ngc1052-mwl,osorioClavijo+2020-ngc1052,falocco+2020-ngc1052}. We first add a \texttt{mekal} component \citep{mewe+1995-mekal} to represent the plasma emission as in the more detailed studies mentioned above, which lowers the total $\chi^2$ to 1491.9. Then, we add an unresolved Gaussian at rest-frame 6.4\,keV, further lowering the total $\chi^2$ to 1322.1. Additional free parameters are the temperature and normalization of the \texttt{mekal} component and the normalization of the emission line. Residuals are again shown in Figure~\ref{fig:sepchis}. Temporarily letting the line energy be a free parameter, we find that it is constrained to $(6.39\pm0.01)$\,keV, identifying it as the neutral \feka\ line. Its equivalent width is $(100\pm10)$\,eV, which is consistent with results from \citet{brenneman+2009-ngc1052} and only marginally lower than the \feka\ equivalent widths from \citet{gonzalezMartin+2009}, \citet{rivers+2013}, and \citet{falocco+2020-ngc1052}.

In order to allow for greater flexibility in the absorption profile, we replace the neutral absorption component with an ionized one represented by the \texttt{zxipcf} model. The additive components of the \xspec\ model are \texttt{zxipcf}$\times$\texttt{cabs}$\times$\texttt{cutoffpl}, \fsca$\times$\texttt{cutoffpl}, \texttt{mekal}, and \texttt{zgauss}. We keep the partial covering fraction fixed at unity, therefore adding only one new free parameter, the ionization parameter, $\xi$ (defined as $L/nR^2$, where $L$ is the ionizing radiation luminosity, $n$ is the gas density, and $R$ is its distance from the radiation source). In the remainder of this paper, we use \nhlos\ to refer to the column density of the partially ionized material in the line of sight. This is the simplest model fitting the data well, yielding $\chi^2/\nu=1253.2/1196=1.048$. As this corresponds to \pnull$=12$\,\%, above our 5\,\% threshold for rejecting a model, this model represents a satisfactory description of the \xmmnewton\ and \nustar\ data. A similarly good fit can be achieved with a partial covering neutral absorption model, but we favor the ionized absorption model because it is easier to interpret. The residuals of this model (second panel from the bottom in Figure~\ref{fig:sepchis}) do not show any further structure except for possible narrow line-like features at some energies (e.g., around 1.9\,keV), which we do not include for simplicity because the model is already statistically acceptable.

The goal of our study is to constrain the properties of the circumnuclear material using its line-of-sight and globally averaged column density. We therefore employ a model to self-consistently represent X-ray reprocessing (Compton scattering, fluorescence, and absorption) in a torus-like geometry. Replacing the phenomenological Gaussian at 6.4\,keV, we include the table model \borustwo\ \citep{balokovic+2018}, which includes both continuum and line emission from reprocessing in a neutral medium characterized by a covering factor (\cftor) and a column density averaged over all covered lines of sight (\nhtor).\footnote{Specifically, we use the FITS table \texttt{borus02\_v170323a.fits}.} We link the parameters related to the intrinsic spectrum to those of existing components, fix the relative Fe abundance to unity (i.e., Solar value), and assume the viewing angle constrained by measurements of the twin jets in the radio band: \thetainc$=80^{\circ}$. The resulting model has only one additional free parameter, as \cftor\ and \nhtor\ are added to the pool and the normalization of the \feka\ line is eliminated. The model fits the data only slightly better than the previous model ($\chi^2/\nu=1247.9/1195=1.044$, \pnull$=14$\,\%); however, its added value is the ability to directly constrain some of the basic properties of the circumnuclear material from the X-ray data.

The best-fit parameters for this model, which we adopt as the basis of our multi-epoch spectral analysis and refer to as model~S hereafter, are given in Table~\ref{tab:models}. In Figure~\ref{fig:eemod} we show the model and highlight its components individually. The observed 2--10\,keV flux calculated from this model is $(5.8\pm0.2)\times10^{-12}$\,\fluxunit, very close to the average of $5.9\times10^{-12}$\,\fluxunit\ observed with \rxte\ (see Figure~\ref{fig:lcs}). The intrinsic luminosity in the same band is $(6.5\pm0.3)\times10^{41}$\,erg\,s$^{-1}$. Noting that \nhlos\ and \nhtor\ converge toward similar values ($\simeq1.3\times10^{23}$\,\cmmt), we also test the assumption that they are equal. This adds self-consistency to the model as the \borustwo\ component formally represents a torus of uniform density. The additional constraint degrades the fit quality negligibly ($\chi^2/\nu=1248.9/1196=1.044$, \pnull$=14$\,\%) and only marginally changes the constraint on \cftor\ from $>\!80$\,\% to $(70\pm10)$\,\%, without shifting any other spectral parameters outside of their 68\,\% confidence intervals. The resulting constraint on \nhlos=\nhtor\,$=\left(1.5_{-0.4}^{+0.2}\right)\times10^{23}$\,\cmmt\ is very close to the value of \nhlos\ in model~S and only marginally different from \nhtor\ given the derived 68\,\% confidence interval. This is consistent with expectations from the small difference in $\chi^2$ and the fact that this constraint is driven by a combination of several spectral features.

\subsection{Joint Multi-epoch Spectral Modeling} 
\label{sec:fitting-multiepoch} 

\begin{figure*}
\begin{center}
\includegraphics[width=\textwidth]{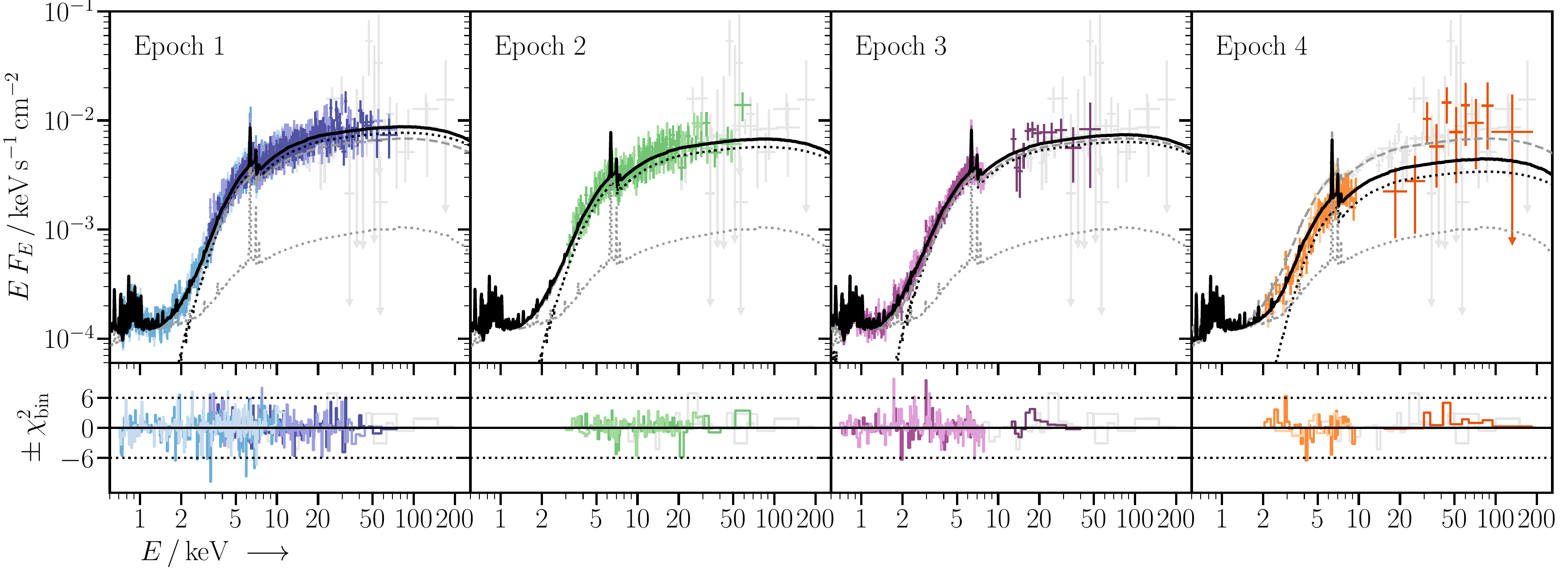}
\vspace{-0.5cm}
\caption{ Unfolded spectra and the best-fit spectral model~B (free \nhtor, variable $\xi$) in the top panels with $\chi^2$ contributions for each energy bin in the bottom panels. The colors of data points match those shown in Figure~\ref{fig:counts}, except the long-term averaged spectra (from \swift, \integral, and \rxte) plotted in the same shade of light grey in the background of each panel. Downward arrows mark error bars formally extending to zero. Thick black lines in each panel show the total spectral model per epoch, and the thick, grey, dashed line shows the total for the average spectrum. Dotted black lines show the line-of-sight component, which is different in every epoch, while the grey dotted lines show the sum of all non-variable components (reprocessing by the torus, secondary power-law component, and plasma emission). \label{fig:eeuf-b}}
\end{center}
\end{figure*}

Before starting our multi-epoch spectral analysis, we verified that each of the four epochs (defined in Table~\ref{tab:obslog}) can be fitted well with the model based on the highest-quality epoch analysed in detail in the preceding section. Consistent with previous studies, we find that the spectrum of NGC\,1052 is qualitatively similar in other epochs, with the differences fully accounted for by variations in the luminosity of the intrinsic continuum, \nhlos, and $\Gamma$. For practical reasons discussed in \S\,\ref{sec:discussion-variability}, in this study we do not consider $\Gamma$ to be variable between epochs. We note that \xmmnewton\ and \suzaku\ spectra at the lowest energies differ, so we ignored spectra from both of them below 0.7\,keV for the joint spectral analysis since the soft X-ray emission is not the focus of our study. The difference may be due to the much smaller extraction region size used for \xmmnewton\ combined with PSF correction calculated assuming only a point source, while the soft emission is extended over the central $\simeq$30\arcsec\ in \chandra\ images analyzed by \citet{osorioClavijo+2020-ngc1052} and \citet{falocco+2020-ngc1052}.

The basic principle behind our joint multi-epoch analysis is the idea that spectra integrated over long periods of time (as in the case of \swiftbat, \integral/ISGRI, and \rxte/PCA) be well described by a model consisting of components constant over long periods and averages of components that vary between epochs. Examples of the former are the extended plasma emission, Thomson-scattered continuum typically associated with the ionization cones \citep[i.e., the narrow-line region; e.g.,][]{gupta+2021}, and reprocessing in the obscuring torus. The large physical extent of the sources of these components (pc--kpc) makes them insensitive to short-timescale variations of the intrinsic continuum. We therefore define a multi-epoch \xspec\ model whose parameters have values that are either equal in all epochs (i.e., treated as a single free parameter) or different in each epoch (i.e., treated as a free parameter in each epoch). In the latter case, the model for the long-term average spectrum is defined by parameter values that are averages of values from individual epochs. In a sense, the long-term spectrum can be seen as an additional, special epoch.

\begin{deluxetable*}{lccccc}
\tablecaption{Parameters of several representative single-epoch and multi-epoch broadband X-ray spectral models \label{tab:models}}
\tabletypesize{\small}
\tablehead{
  \colhead{Parameter\,\tablenotemark{a}} &
  \colhead{Epoch\,\tablenotemark{b}} &
  \colhead{Model~S} &
  \colhead{Model~A\arcmin} &
  \colhead{Model~B} &
  \colhead{Model~B$_3$}
}
\startdata
$\chi^2\,/\,\nu$ &  & 1247.9/1195 & 1912.2\,/\,1836 & 1899.3\,/\,1832 & 1790.7\,/\,1755\\
\pnull\,/\,\% &  & 14 & 11 & 13 & 27 \\
\hline
$\Gamma$ &  & $1.76\pm0.03$ & $1.74\pm0.02$ & $1.72\pm0.02$ & $1.76\pm0.01$ \\
\fsca\,/\,\% & & $3.2\pm0.1$ & $4.2\pm0.2$ & $4.4\pm0.2$ & $3.9_{-0.2}^{+0.1}$\\
\nhtor\,/\,$10^{23}$\,\cmmt &  & $1.01_{-0.08}^{+0.06}$ & =\nhlos\,(0) & $1.2\pm0.1$ & $1.7_{-0.2}^{+0.1}$\\
\cftor\,/\,\%\,\tablenotemark{c} &  & $100_{-20}^{+u}$ & $80\pm4$ & $100_{-10}^{+u}$ & $80\pm4$\\
\hline
\multirow{5}{*}{$K$\,/\,$10^{-3}$\,keV$^{-1}$\,s$^{-1}$\,cm$^{-2}$}
 & 0 & \nodata & $(2.7\pm0.1)$ & $(2.6\pm0.1)$ & $(3.11\pm0.08)$ \\
 & 1 & $3.6\pm0.1$ & $3.6\pm0.2$ & $3.5\pm0.2$ & $3.6\pm0.1$ \\
 & 2 & \nodata & $2.6\pm0.1$ & $2.5\pm0.2$ & $2.7\pm0.1$\\
 & 3 & \nodata & $2.9\pm0.1$ & $2.9\pm0.1$ & $3.0\pm0.1$ \\
 & 4 & \nodata & $1.6\pm0.4$ & $1.5\pm0.3$ & \nodata \\
\hline
\multirow{5}{*}{\nhlos\,/\,$10^{23}$\,\cmmt}
 & 0 & \nodata & $(1.83\pm0.04)$ & $(1.84\pm0.05)$ & $(1.66\pm0.06)$ \\
 & 1 & $1.6_{-0.1}^{+0.2}$ & $1.92\pm0.06$ & $2.00\pm0.04$ & $1.61_{-0.06}^{+0.03}$ \\
 & 2 & \nodata & $1.8\pm0.1$ & $1.8\pm0.3$ & $1.7\pm0.1$ \\
 & 3 & \nodata & $1.64\pm0.06$ & $1.68\pm0.06$ & $1.65\pm0.04$ \\
 & 4 & \nodata & $2.0\pm0.1$ & $1.9\pm0.2$ & \nodata \\
\hline
\multirow{5}{*}{$\log\,(\,\xi$\,/\,erg\,s$^{-1}$\,cm$^{-1}\,)$}
 & 0 & \nodata & $1.43\pm0.08$ & $(1.3\pm0.1)$ & $(1.26\pm0.08)$ \\
 & 1 & $1.09_{-0.04}^{+0.01}$ & $=\log\,\xi$\,(0) & $1.53\pm0.08$ & $1.0\pm0.1$\\
 & 2 & \nodata & $=\log\,\xi$\,(0) & $1.4\pm0.3$ & $1.3_{-0.1}^{+0.2}$ \\
 & 3 & \nodata & $=\log\,\xi$\,(0) & $1.45\pm0.09$ & $1.41\pm0.04$ \\
 & 4 & \nodata & $=\log\,\xi$\,(0) & $0.8\pm0.4$ & \nodata \\
\enddata
\tablenotetext{a}{Two \texttt{mekal} parameters that do not change value between the models are not listed in the table: its temperature, $0.68\pm0.03$\,keV, and normalization factor, $(2.4\pm0.2)\times10^{-5}$. Fitted cross-normalization factors are listed separately in Table~\ref{tab:cnfactors}.}
\tablenotetext{b}{Epoch index zero marks the long-term average values. Cases in which a parameter is not fitted but calculated as the average of individual epochs are given in parentheses.}
\tablenotetext{c}{$+u$ is given when the uncertainty is consistent with the upper end of the parameter domain.}
\end{deluxetable*}

For our simplest multi-epoch model we choose to fit for single, shared values of spectral parameters $\Gamma$, \fsca, $\xi$, \cftor, and \nhtor, in addition to the \texttt{mekal} parameter $kT$ and its normalization. Parameters \nhlos\ and $K$ (intrinsic continuum normalization) are free parameters in each of the four epochs, while the corresponding parameters for the average ``epoch'' are set to be the averages over the other four epochs. We also fit for 10 cross-normalization factors ($C_{\rm \scriptsize inst}$) while assuming that one spectrum per epoch has this factor fixed at unity. These factors can be arranged in a number of different ways, which we discuss in more detail in the Appendix. For this model, which we call model~A, the $4\!+\!1$-epoch fit has $\nu=1835$. It provides a good fit to the data with $\chi^2=1904.5$ (\pnull$=13$\,\%). Again noting the convergence of the average \nhlos\ and \nhtor\ ($\simeq\!1.5\times10^{23}$\,\cmmt), we test the assumption that they are equal. Model~A\arcmin, which includes this additional constraint, yields $\chi^2=1912.2/1836=1.041$ (\pnull$=11$\,\%) for the parameter values listed in Table~\ref{tab:models} with their uncertainties representing 68\,\% confidence intervals. The only notable difference in parameters compared to model~A is \cftor$\,=(80\pm4)$\,\% as opposed to $>\!90$\,\%, while all other parameters stay within their derived uncertainties.

We try to improve the model further by additionally letting the parameter $\xi$ have a different value in each epoch, like \nhlos\ and $K$ in model~A. This model (named~B) has three additional free parameters and fits the data slightly better: $\chi^2=1899.3/1832=1.037$ (\pnull$=13$\,\%). The same is true for the primed version of the model (B\arcmin) with \nhtor\ equal to \nhlos\ averaged over epochs ($\chi^2=1902.5/1833=1.038$, \pnull$=13$\,\%). Since the parameter constraints are very similar, we list them only for model~B in Table~\ref{tab:models}. We also show this model in Figure~\ref{fig:eeuf-b} as an illustrative example for all models mentioned in this section, because they are too similar to visually distinguish any differences. The slight differences produced by $\xi$ variations or the average---as opposed to independently fitted---\nhtor\ are negligible, as expected from the very small difference in the total $\chi^2$ between these well-fitting models. Given the very small differences in the best-fit reduced $\chi^2$ and \pnull\,$>5$\,\% in all cases, it is not possible to formally select the preferred scenario on statistical grounds.

Although we considered a number of other possible improvements and alternatives to models A, A\arcmin, B, and B\arcmin, we did not find any that resulted in a significant decrease of the total $\chi^2$. We note, however, that the spectrum in epoch~4 (\bepposax\ data) seems to be most different from the other epochs, with the lowest intrinsic continuum normalization and the lowest ionization parameter. As an additional test, we performed the multi-epoch analysis excluding \bepposax\ data. We find the biggest differences for model~B$_3$ (the 3-epoch equivalent of model~B), which we include in Table~\ref{tab:models} for direct comparison. Consistent with expectations, the average normalization of the continuum is higher, but the spectral parameters do not generally change appreciably. The data considered in this spectral analysis may not allow us to distinguish between the slightly different models described above, but they all point toward a self-consistent multi-epoch solution for the broadband X-ray spectrum of NGC\,1052 with interesting physical constraints further discussed in \S\,\ref{sec:discussion-torus}.

\subsection{2--10\,keV Light Curve Modeling} 
\label{sec:fitting-lc} 

\begin{figure}
\begin{center}
\includegraphics[width=\columnwidth]{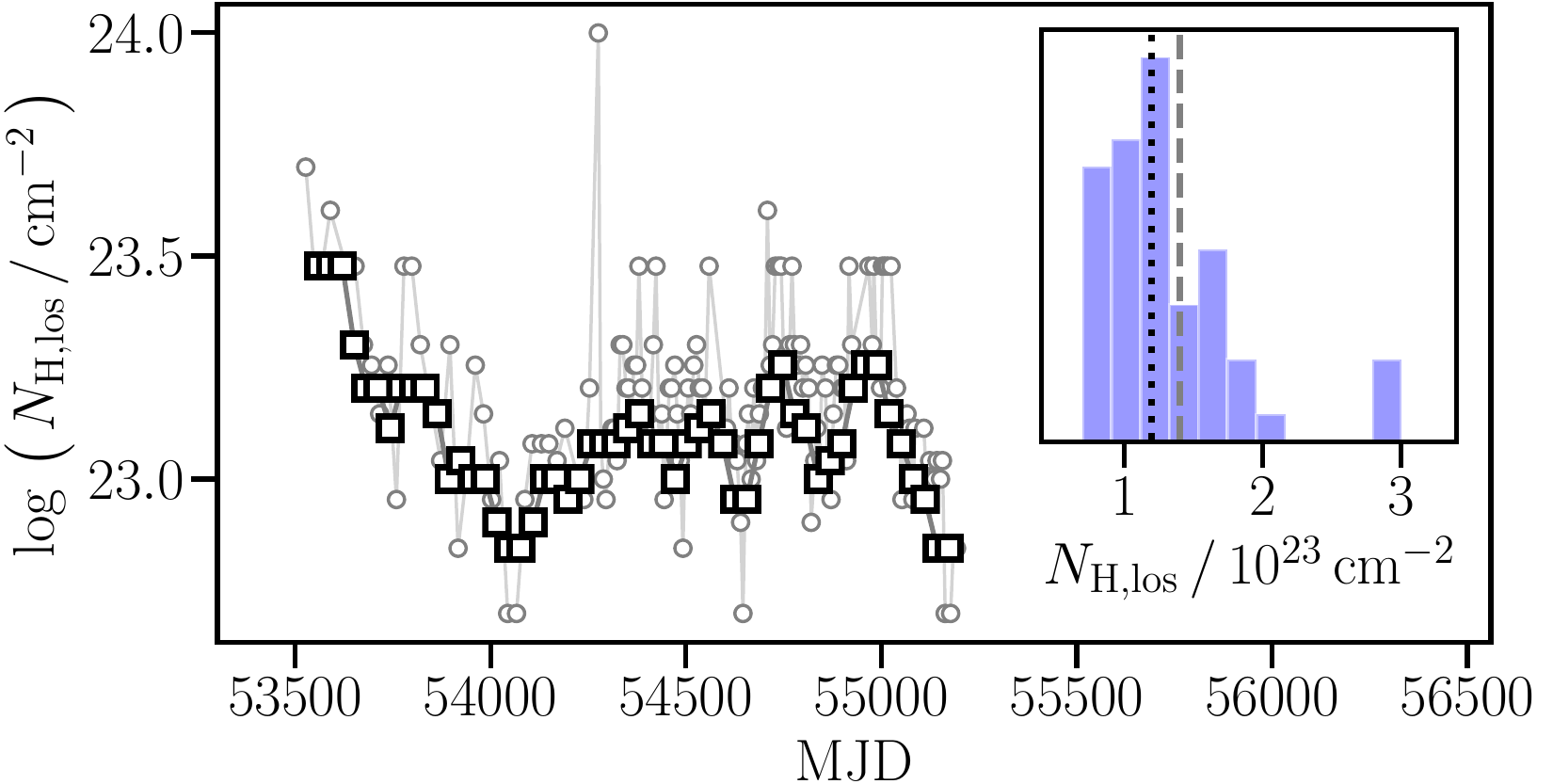}
\vspace{-0.4cm}
\caption{ Inferred variability of the line-of-sight column density (\nhlos) over the time period covered by \rxte/PCA observations. \nhlos\ values were obtained from fitting the observed fluxes in the 2--4\,keV, 4--7\,keV, and 4--10\,keV bands shown in Figure~\ref{fig:lcs}; see \S\,\ref{sec:fitting-lc} for details. Grey symbols and lines show values from original single observations, while the black ones show those obtained from three-month averages. In the inset we show the histogram for the three-month averaged \nhlos, and medians obtained from averaged (black dotted line) and original (grey dashed line) data. \label{fig:nhlc}}
\end{center}
\end{figure}

The \rxte/PCA light curves in the 2--10\,keV band shown in Figure~\ref{fig:lcs} (bottom panel) offer an additional self-consistency test for our broadband X-ray spectral model for NGC\,1052. We first employ models~A and A\arcsec\ for the average spectrum from \S\,\ref{sec:fitting-multiepoch} to calculate observable fluxes in the 2--4\,keV, 4--7\,keV, and 7--10\,keV bands as a function of \nhlos\ for the range $22<$\lognhloscmmt$<24$. Then, for each time bin with flux measured in all three bands, we calculate $\chi^2$ as a function of \nhlos\ and identify the lowest value. With three points per time bin, one free parameter per bin, and 148 (56) bins in the original (3-month averaged) light curves, the total number of degrees of freedom in this fitting problem is 296 (112). While the exact value of $\chi^2$ depends on the assumed systematic uncertainties mentioned in \S\,\ref{sec:obs-longterm}, we are able to find overall good fits with total $\chi^2$ below about 350 (150) even assuming the minimal level of systematics. The resulting variability in \nhlos\ during the 4.5-year period covered by \rxte\ observations is shown in Figure~\ref{fig:nhlc}. Identical results follow from assuming model~A\arcmin\ instead of~A. The main outcome of this analysis is the inferred distribution of \nhlos, which would be challenging to obtain directly from time-resolved spectral fitting of \rxte\ data.

As shown in Figure~\ref{fig:nhlc} (inset), \nhlos\ is distributed in the range between $8\times10^{22}$\,\cmmt\ and $2\times10^{23}$\,\cmmt, with a few outliers around $3\times10^{23}$\,\cmmt\ mostly from observations in late 2005. The medians for the original and three-month averaged light curves are very close: $1.2\times10^{23}$\,\cmmt\ and $1.4\times10^{23}$\,\cmmt, respectively. The central 68\,\% of the distribution is between $0.9\times10^{23}$\,\cmmt\ and $1.7\times10^{23}$\,\cmmt. For this calculation we assumed that the intrinsic continuum does not vary, which can be justified by the absence of any notable trend in the \swiftbat\ light curve (Figure~\ref{fig:lcs}, top panel) over the same time period. If we allow for a small vertical offset and a linear trend in intrinsic luminosity, we get even lower total $\chi^2$, but the \nhlos\ distribution is not significantly affected. However, this is likely over-fitting the available data. Likewise, fitting a grid of models with variability in both \nhlos\ and $\xi$ (as in models~B and~B\arcmin) provides too much freedom for the data considered here, although from manual comparison for a subset of time bins, we typically find \nhlos\ slightly above $1\times10^{23}$\,\cmmt.

\section{Results and Discussion} 
\label{sec:discussion} 

\subsection{Broadband X-ray Spectrum} 
\label{sec:discussion-broadband} 

Our analysis establishes a broadband X-ray spectral model for NGC\,1052 that is uniquely based on essentially all currently available hard X-ray data and self-consistently accounts for observed variability over several epochs sparsely covering 17 years. The exquisite broadband coverage is particularly helpful for better constraining the parameters of broad continuum components. Apart from the novel aspect of the physically motivated reprocessed component represented by the \borustwo\ model, the main features of our models are similar to those of models in the literature. Our modeling is deliberately not exhaustive; it is rather focused on demonstrating the advantages of our approach for interpretation of X-ray data in terms of physical rather than phenomenological parameters.

Prior to the long \suzaku\ observation in 2007 \citep{brenneman+2009-ngc1052} the absorption profile was not considered to be due to partially ionized gas, which typically led to inferences of $\Gamma<1.5$ for the intrinsic continuum \citep{weaver+1999-ngc1052,guainazzi+2000-ngc1052,kadler+2004a-ngc1052-mwl}. Detailed studies by \citet{brenneman+2009-ngc1052}, \citet{osorioClavijo+2020-ngc1052}, and \citet{falocco+2020-ngc1052} considered both ionized and multi-layer, partial-covering absorption reporting $\Gamma$ in the wide range 1.2--1.7 depending on the epoch and the assumed spectral model. \citet{rivers+2013} and \citet{cabral-2020} found $\Gamma$ consistent with 1.65, close to the typical range observed in bright, nearby Seyferts selected in the hard X-ray band \citep{balokovic-2017,ricci+2017-bass,panagiotou+walter-2019}. Our constraints cluster around $\Gamma=1.74$, with statistical uncertainties approximately equal to systematic uncertainties due to model selection (each contributing about $\pm0.02$). This is under the assumption that  \ecut\,$=290$\,keV, based on the current best estimate for the median in the nearby obscured AGN population \citep{balokovic+2020}.

Our data and models provide the most robust direct constraint on \ecut\ in NGC\,1052 to date. Letting \ecut\ be a free parameter in the multi-epoch fits for models considered in \S\,\ref{sec:fitting-multiepoch}, we uniformly find that the best-fit value tends to be at the high-energy end of the parameter domain. The models also agree that a lower limit on \ecut\ is around 220\,keV at the 68\,\% confidence level and around 140\,keV at the 99\,\% level. As expected from \ecut\ in this energy range, our constraint on $\Gamma$ does not change with respect to fixed \ecut, but uncertainties for each model increase to typically $\pm0.04$. Comparing to $\Gamma=1.36\pm0.09$ and \ecut$=80_{-20}^{+40}$\,keV from \citet{balokovic+2020}, based on poorer-quality data and a simpler spectral model, we further support their claim that such low and apparently well-constrained cutoffs are likely just a consequence of degeneracy between model parameters. We also make use of the recently published equivalent of the \borustwo\ model named \texttt{borus12} \citep{balokovic+2019-rnaas}, which features a more physical Comptonized continuum (\texttt{nthcomp}; \citealt{zdziarski+1996,zycki+1999}) in place of the phenomenological cutoff power law.\footnote{Specifically, we use the FITS table \texttt{borus12\_v190815a.fits}.} Instead of \ecut, models with this intrinsic continuum directly provide a lower limit on the electron temperature of the corona: $kT_e>120$\,keV at the 68\,\% confidence level and $kT_e>50$\,keV at the 99\,\% level.

One additional feature that previous studies considered is the contribution from relativistic disk reprocessing (i.e., relativistic reflection). Although \citet{brenneman+2009-ngc1052} found tentative evidence for relativistic reflection around the \feka\ line in \suzaku\ data, more recent studies---including this one---did not find it necessary to include such a component in order to fit the observed spectra sufficiently well. We attempted to force a relativistically broadened reflection component into our already well-fitting models by manually increasing its normalization from close to zero. For this exercise we used the \texttt{relxill} model \citep{garcia+2014-relxill,dauser+2014-relxill}, which is based on the same intrinsic continuum as \texttt{borus12}. All shared parameters were linked, and we fixed parameters relevant for innermost disk reflection to the ranges of values discussed in \citet{brenneman+2009-ngc1052} and \citet{falocco+2020-ngc1052}. However, we were unable to find any configuration in which $\chi^2/\nu$ decreased significantly with the addition of this component. Its normalization generally drifted toward zero in the calculation of uncertainties in \xspec. We therefore conclude that we find no evidence for relativistic reflection in the X-ray spectrum of NGC\,1052.

\subsection{Variabilty} 
\label{sec:discussion-variability} 

Variable intrinsic X-ray luminosity is commonly observed in AGN. In our analysis we allow the intrinsic luminosity to be different in each of the four epochs with the constraint that the average has to match the intrinsic luminosity of the spectra from the BAT, ISGRI, and PCA instruments, which are averaged over long time periods ($>\!4$ years in each case). We find that over the four epochs the intrinsic luminosity varies with an amplitude of at most 30\,\% around the mean of $4.7\times10^{41}$\,erg\,s$^{-1}$ in the 2--10\,keV band. This is consistent with the relatively stable flux in the BAT band (Figure~\ref{fig:lcs}, top panel), which is dominated by the intrinsic continuum.

Marginal evidence for the scaling of $\Gamma$ with luminosity over its modest variability amplitude has previously been found by \citet{hernandezGarcia+2013-liners} and \citet{connolly+2016}, but in the opposite sense. In our analysis we assumed that $\Gamma$ is not variable because the normalizations of the \texttt{cutoffpl} and \borustwo\ components are highly correlated with $\Gamma$. Calculation of the average intrinsic continuum and therefore the normalizations of the non-variable components in \xspec, which is a key ingredient of our multi-epoch modeling method, is formally correct only under this assumption. Constant $\Gamma$ seems to be statistically consistent with the data considered in our analysis. However, a different setup of the model in future work would make it possible to test the $\Gamma-$\lamedd\ relationship in the range of \lamedd\ where the scaling appears to reverse sign \citep{constantin+2009,connolly+2016,she+2018}.

Variability of \nhlos\ in NGC\,1052 has been well established in previous studies (e.g., \citealt{hernandezGarcia+2013-liners,falocco+2020-ngc1052,osorioClavijo+2020-ngc1052}). A detailed comparison of \nhlos\ values is hampered by the differences in adopted spectral models and the simultaneous fitting for $\Gamma$ in each epoch leading to the well-known degeneracy between these two spectral parameters. Our multi-epoch spectral modeling and modeling of the \rxte\ light curves are generally consistent with findings in the literature, which fall within the range $7\times10^{22}<$\,\nhlos\,/\,\cmmt\,$<2.5\times10^{23}$ for the highest-absorption components. From spectroscopy, we find that \nhlos\ varies by no more than 20\,\% around the average which is $(1.7-1.8)\times10^{23}$\,\cmmt, depending on the model (see Table~\ref{tab:models}). \rxte\ light curves imply a larger amplitude of variability---up to a factor of two---around a slightly lower median, $(1.2-1.4)\times10^{23}$\,\cmmt. Light curve modeling may not be as reliable as full spectral modeling, but still represents a valuable consistency test. The most recent broadband X-ray spectral modeling results from \citet{osorioClavijo+2020-ngc1052} and \citet{cabral-2020}, based on a larger number of epochs but fewer hard X-ray spectra, suggest variability in \nhlos\ similar to the range presented here.

\subsection{Circumnuclear X-ray Reprocessing} 
\label{sec:discussion-torus} 

Most previous studies of NGC\,1052 broadband X-ray spectra \citep[e.g.,][]{brenneman+2009-ngc1052,balokovic-2017,osorioClavijo+2020-ngc1052} included a spectral component accounting for reprocessing in circumnuclear material are based on the simplifying assumption that the Compton hump is largely fixed in shape and well represented by X-ray reprocessing models with the geometry of a slab of infinite column density (e.g., \pexrav; \citealt{magdziarz+zdziarski-1995}). The use of models similar to \pexrav\ is broadly justified across the literature by the expectation that the circumnuclear material obscuring the central source of X-rays in AGN exists in the form of a torus-like structure containing gas and dust with CT column densities. However, the shape of the reprocessed continuum depends on the geometry and the column density of the circumnuclear material. In Figure~\ref{fig:eemod} we contrast the difference in the spectral shape of the best-fit reprocessed component from our single-epoch modeling (\S\,\ref{sec:fitting-singleepoch}) with that of the \pexrav\ model. Its normalization ($\vert\,$\rpex$\,\vert <0.3$) is set according to the upper limit based on the most directly constraining \nustar\ data from \citet{balokovic-2017}, which is consistent with $\vert\,$\rpex$\,\vert \approx0.1$ from additional \nustar\ data considered in \citet{osorioClavijo+2020-ngc1052}.

\begin{figure}
\begin{center}
\includegraphics[width=\columnwidth]{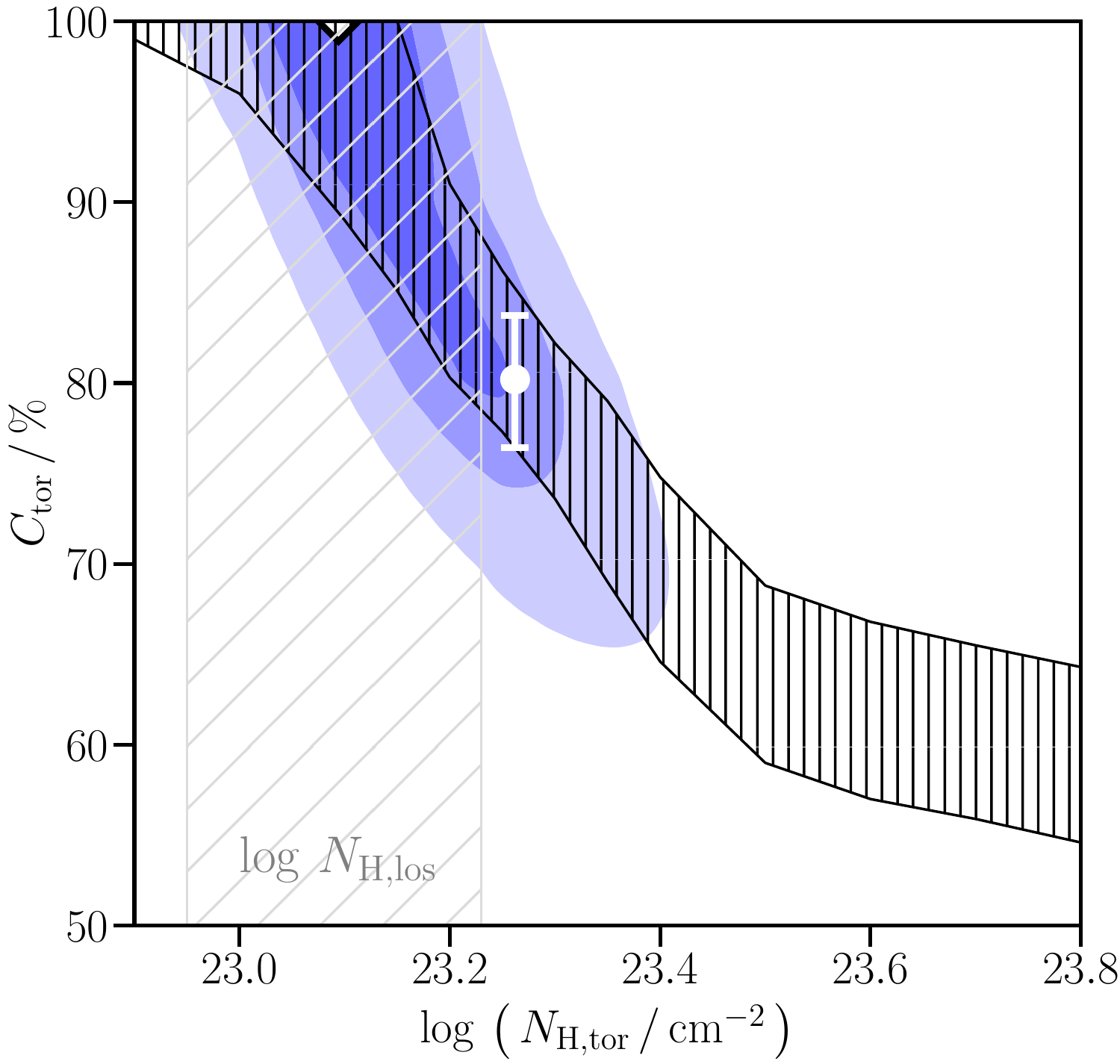}
\vspace{-0.6cm}
\caption{ Constraints on the main torus parameters (covering factor, \cftor, and column density, \nhtor) based on models~A and~A\arcmin. Blue contours show 1\,$\sigma$, 2\,$\sigma$, and 3\,$\sigma$ confidence regions going from darker to lighter colors, respectively. The open black triangle at the upper end of the \cftor\ range marks the best fit for model~A, in which \nhtor\ is a free parameter. The band covered with black vertical hatching shows 1\,$\sigma$ confidence region for fits in which \nhtor\ was fixed at values within the plotted range. The white circle and its error bar show the \cftor\ constraint from model~A\arcmin, in which \nhtor\ is assumed to be equal to the average \nhlos. The same type of plot for models~B and~B\arcmin, as well as three-epoch fits, would look very similar. The light grey hatched area shows the range of \nhlos\ recovered from our analysis of the \rxte\ light curves. Both the multi-epoch spectral analysis and the light curve modeling suggest that \nhlos\,$=(1-2)\times10^{23}$\,\cmmt, overlapping with our constraints on the \nhtor\ parameter. \label{fig:contsweep}}
\end{center}
\end{figure}

Our analysis, based on a model in which the reprocessed component (both the continuum and lines treated self-consistently) is represented by the \borustwo\ model, finds that the X-ray data are consistent with a nearly spherical distribution of circumnuclear material with a high covering factor (80--100\,\%) and column density significantly below the CT threshold ($1-3\times10^{23}$\,\cmmt). In this regime, the Compton hump is significantly lower and broader (i.e., less strongly peaked) than in \pexrav\ and similar models, as Figure~\ref{fig:eemod} demonstrates. These properties provide a straightforward explanation for the previous measurements, but seem to differ from the expectations for the classical AGN {\em torus}. Despite the common nomenclature, which we keep, neither obscuration nor reprocessing in AGN are necessarily tied to the observationally established dusty torus. Direct measurements of the properties of X-ray reprocessing structures have only recently become feasible, and studies like the one presented here are only beginning to test the correspondence \citep{esparzaArredondo+2019-ic5063,esparzaArredondo+2021,ogawa+2021}.

Both the multi-epoch spectral analysis and the results from light curve modeling suggest that \nhlos\ is $(1-3)\times10^{23}$\,\cmmt, overlapping with our constraints on the \nhtor\ parameter of the \borustwo\ component. We show this visually in Figure~\ref{fig:contsweep} for models~A and A\arcmin, noting that models~B and~B\arcmin\ provide almost identical constraints. The average \nhlos\ for four individual epochs, ($1.83\pm0.04)\times10^{23}$\,\cmmt, is just slightly above the range containing 68\,\% of the \nhlos\ distribution determined from \rxte\ light curves and shown with the grey hatched region in Figure~\ref{fig:contsweep}. Excluding the epoch 4 data lowers the average, $(1.66\pm0.06)\times10^{23}$\,\cmmt, to within this interval. Taken together at face value, these results could be interpreted as due to an atypical, nearly spherical, roughly uniform, low-density ``torus'' responsible for both reprocessing and obscuration.

To test this interpretation, we perform a series of fits to the multi-epoch spectral data with the parameter \nhtor\ fixed at a range of values in the range $22.5<$\,\lognhtorcmmt\,$<24.5$ for both A and B models. With the black hatched region in Figure~\ref{fig:contsweep} we show the constraints on the parameter \cftor\ in the former case. The range of the horizontal axis of the figure is set by the \nhtor\ interval within which the fits result in a $\chi^2$ low enough that the models cannot be rejected (i.e., \pnull\ exceeds 5\,\%); outside of the plotted range models are no longer acceptable. From this exercise we conclude that while a torus with \lognhtorcmmt$\,\approx23.5$ and \cftor$\,\approx65$\,\% is still statistically consistent with the data, one with \lognhtorcmmt$\,\gtrsim24$ is not. However, this is a model-dependent statement, and an alternative model with a non-uniform density distribution may contain a small fraction of lines of sight covered with CT column densities as long as the global average is closer to the observed \nhlos.

In order to examine one possible alternative, we employed the \borus\ code \citep{balokovic-2017,balokovic+2018} to compute a new table model for \xspec\ assuming a spherical geometry and a vertical density gradient, such that $n\propto 10^{-z}$, where $z$ is the coordinate perpendicular to the equatorial plane of the torus. We quantify the gradient with the parameters \nhequ, the equatorial column density, and \dctor, the density contrast between the equatorial plane and the height equal to the outer radius of the torus above and below that plane. We cover the parameter space, $1\leq$\,\logdctor$\leq5$ and $22.5\leq$\,\lognhequcmmt$\leq24.5$, with 5 equidistant points in each direction. The case \logdctor$=0$ corresponds to uniform density, which we recover from \borustwo\ with \cftor\,$=100$\,\%. In the limited version used here, \nhequ\ and \dctor\ simply replace \nhtor\ and \cftor\ parameters of \borustwo.We note that this model, like \borustwo, incorporates only absorption and fluorescence in neutral material. We verified in \S\,\ref{sec:fitting-singleepoch} that the prominent \feka\ line corresponds to largely neutral reprocessing. Accounting for mildly ionized material could at most only marginally alter the reprocessed component at $\lesssim5$\,keV, where the absorption profile becomes relevant (see Figure~\ref{fig:eemod}). The full model with expanded parameter space will be described in a future publication.

\begin{figure}
\begin{center}
\includegraphics[width=\columnwidth]{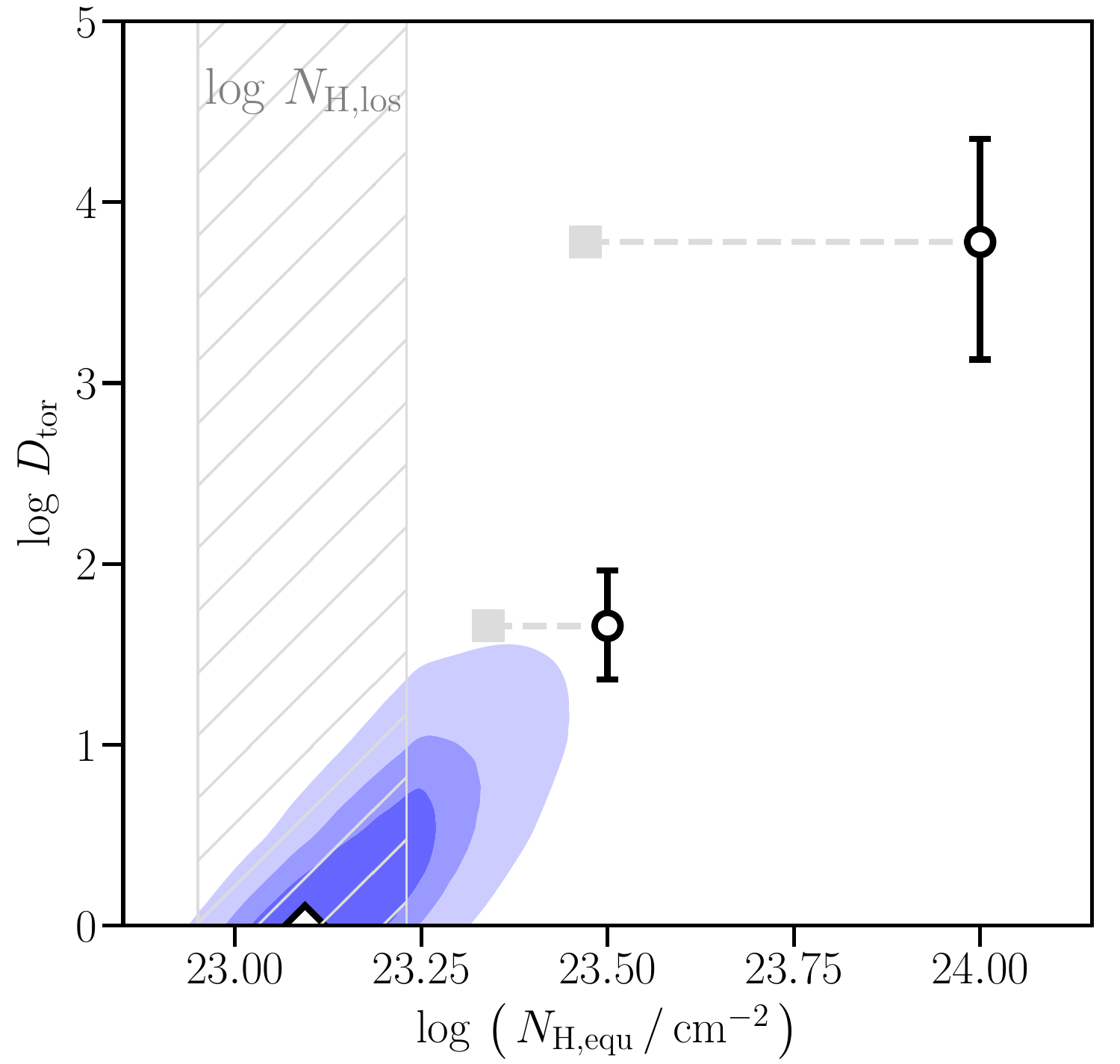}
\vspace{-0.6cm}
\caption{ Constraints on the parameters of the torus model featuring a vertical density gradient, parameterized by the equatorial column density, \nhequ, and the density contrast parameter, \dctor. Blue contours show 1\,$\sigma$, 2\,$\sigma$, and 3\,$\sigma$ confidence regions going from darker to lighter colors, respectively. The open black triangle marks the best fit, which is consistent with uniform-density torus. Black open circles and their error bars show constraints on \dctor\ at two fixed values of \nhequ\ that are more than 3\,$\sigma$ away from the best fit, although they still represent statistically acceptable models. The associated grey squares show the predicted \nhlos\ at the inclination of $80^{\circ}$ for the given pair of \nhequ\ and \dctor\ values. The light grey hatched area shows the range of \nhlos\ recovered from our analysis of the \rxte\ light curves. Within the domain of this spectral model for the torus, it is not possible to self-consistently account for the observed range of \nhlos. \label{fig:borus04}}
\end{center}
\end{figure}

The results of fitting the multi-epoch data with the reprocessing model featuring a vertical density gradient are summarized in Figure~\ref{fig:borus04}. We start with the assumption that \lognhequcmmt\,$=24.0$, finding that the data require a steep density gradient with \logdctor\,$=3.8\pm0.6$. Both the equivalent of model~A and model~B fit the data sufficiently well ($\chi^2/\nu=1930.3/1836=1.051$, \pnull$=6$\,\%, and $\chi^2/\nu=1920.0/1833=1.047$, \pnull$=8$\,\%, respectively), but not better than their uniform-density counterparts. Assuming \lognhequcmmt\,$=23.5$ fits the data slightly better for a lower \dctor, while for a free \nhequ\ the fits converge to the original results presented in \S\,\ref{sec:fitting-multiepoch} (\logdctor\,$=0$) within the confidence regions shown in Figure~\ref{fig:borus04}. In the figure we also show the model-based \nhlos\ at the inclination angle of 80$^{\circ}$ that we assumed for NGC\,1052 throughout our analysis. We fitted \nhlos\ parameters independently here, but these model-based values suggest that, despite the steep density gradient, a torus with \lognhequcmmt\,$\gtrsim23.5$ cannot self-consistently account for the range of \nhlos\ observed in NGC\,1052.

The vertical gradient model belongs to a broader and more flexible class of models than the uniform-density \borustwo. Surprisingly, the results discussed in this section point to the latter as the preferred configuration in terms of the fitting statistic and the ability to account for both reprocessing and line-of-sight obscuration. Within the limitations of our current X-ray model, we interpret this as evidence that NGC\,1052 may be lacking a classical AGN torus with an appreciable covering factor of CT material outside of our line of sight. Without shielding from CT material, it is easier to ionize a significant fraction of the torus, possibly explaining the mildly ionized absorption along our line of sight despite nearly edge-on inclination. Our conclusion might have been different for an inclination of $\lesssim\!45^{\circ}$. However, such inclination is inconsistent with the VLBI observations of the twin relativistic jets in NGC\,1052. Although some misalignment between the torus and jet axes is possible, the most recent constraints on jet orientation suggest an inclination closer to $\simeq80^{\circ}$ \citep{baczko+2016-ngc1052,baczko+2019-ngc1052}. This highlights the importance of taking into account constraints from other wavelengths, which we consider in the following two sections.

\begin{figure*}
\begin{center}
\includegraphics[width=\textwidth]{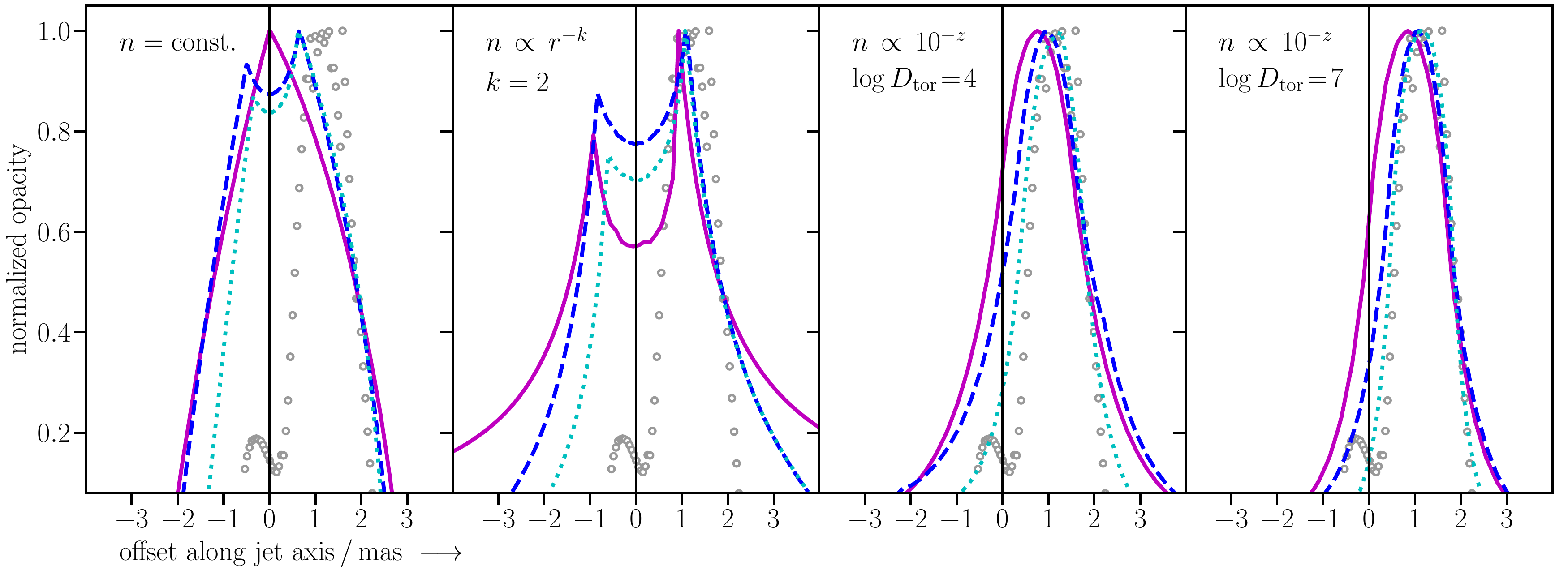}
\vspace{-0.5cm}
\caption{ Normalized opacity as a function of distance along the jet axis for several simple torus models (colored curves) compared to measurements based on VLBI observations of NGC\,1052 jets (\citealt{sawadaSatoh+2008-ngc1052}; grey symbols). The density distribution assumed for the models is given in the upper left corner of each panel, with $r$ and $z$ referring to the radial coordinate and the vertical coordinate perpendicular to the equatorial plane, respectively. In each panel we show model curves for the following cases: model most similar to that fitted to the X-ray data (solid magenta line; see \S\,\ref{sec:discussion-radio} for details), model with a zero-density central cavity with a radius of 25\% of the torus size (blue dashed line), and the same model with an inclination angle of 70$^{\circ}$ instead of 80$^{\circ}$ (cyan dotted line). In all cases, the torus radius is adjusted to match the data around 50\,\% opacity at the outer edge around the offset of $+2$\,mas. \label{fig:radioop}}
\end{center}
\end{figure*}

\subsection{Matching Radio Opacity Measurements} 
\label{sec:discussion-radio} 

In order to compare our torus model to the spatially resolved absorption along the twin jets observed with VLBI in the radio band, we adopt the same approach as \citet{kameno+2001-ngc1052}. For torus geometries discussed in the preceding section, we integrate the optical depth (directly proportional to \nh) as a function of position along the axis defined by the projection of jets onto the sky. We initially assume an inclination of 80$^{\circ}$ as in our X-ray analysis. For simplicity, we ignore the absolute normalization and examine the resulting opacity profiles normalized by their respective maxima. In Figure~\ref{fig:radioop} we compare calculations for a series of models to measurements from \citet{sawadaSatoh+2008-ngc1052}. The zero point of these measurements is the brightest point in the 43\,GHz VLBI image. It corresponds to the dynamic center of the system \citep{vermeulen+2003-ngc1052} and the most recent, detailed studies by \citet{baczko+2019-ngc1052} and \citet{nakahara+2020-ngc1052} confirm that it is consistent with the black hole position to within 0.02\,mas. Positive offsets from the zero point are in the direction of the receding (western) jet. At the distance of 19.5\,Mpc, 1\,mas corresponds to 0.095\,pc.

In the leftmost panel of Figure~\ref{fig:radioop}, we first show that the uniform density model with $80$\,\% covering factor (solid magenta line, exactly corresponding to our \borustwo-based X-ray models) produces a peak at zero offset, unlike the measurements. To create an asymmetry, we test adding a central cavity with zero density extending up to 25\,\% of the radius assumed for the torus (dashed blue line), and we change the inclination to 70$^{\circ}$ (dotted cyan line). As long as the central cavity is small compared to the torus volume, its effect on the reprocessed X-ray spectrum would be negligible. Although these modifications move the opacity profile in the right direction, they are far from a good match with the opacity measurements. This is primarily due to the relatively low contrast between the peak and opacity at zero and negative offsets, i.e., in relative terms, the column density of material in front of the approaching jet is too large in the framework of a uniform-density model.

We can further enhance the asymmetry by assuming a simple radial density profile parameterized as $n\propto r^{-k}$, where $r$ is the radial coordinate. Such a density profile has been assumed for the torus in other models \citep[e.g.,][]{stalevski+2016,fromm+2018}, albeit combined with an anisotropy factor that we test separately. With only the radial dependence, this model is nearly isotropic as seen from the X-ray source at the center, so the difference in X-ray spectra with respect to the uniform model would be minimal. In Figure~\ref{fig:radioop}, second panel from the left, we show a particular case of $k=2$, which is well matched to the results of fitting a complex model to multi-frequency VLBI data for NGC\,1052 by \citet{fromm+2019-ngc1052}. Our conclusions apply more generally to $0.1\!\leq\!k\!\leq\!5$. Again, we find that neither adding a central cavity\footnote{Because of the steep density profile, a small cavity ($\sim\!1$\,\% of the torus radius) is necessary here, otherwise opacity peaks very sharply at zero offset.} nor changing the inclination can create a sufficient opacity contrast to match the measurements. With this family of models it is also difficult to reproduce the sharp drop in opacity at the positive offset of about 2\,mas, which appears to be associated with the outer edge of the torus.

Finally, we examine the model with a vertical density gradient, which we also considered in \S\,\ref{sec:discussion-torus} with respect to the X-ray data. Two cases are shown in the right half of Figure~\ref{fig:radioop}: \logdctor\,$=4$, as a good fit to the X-ray data for the assumption \lognhequcmmt\,$=24$, and a significantly steeper gradient with \logdctor\,$=7$. The latter is out of the parameter domain of our preliminary X-ray model calculation, but matches the radio opacity profile slightly better. The addition of a central cavity and the decrease in inclination (with 25\,\% of the torus radius and 70$^{\circ}$ inclination shown in Figure~\ref{fig:radioop}) bring the opacity profiles even closer to the radio measurements. The qualitative agreement of this model with data in both X-ray and radio bands is an encouraging indication of the direction in which a joint fitting model should be further developed. The steep vertical gradient is also a feature of the \citet{fromm+2018} model, which was found to be consistent with numerous VLBI observations of NGC\,1052 \citep{fromm+2019-ngc1052}.

Within the confines of our current X-ray model, the best fit occurs for \logdctor\,$=0$, but higher values of \logdctor\,$\approx4$ still result in statistically acceptable solutions. We expect that a further increase in the steepness of the density gradient may be able to reproduce the X-ray signature equally well while also concentrating the material toward the equatorial plane of the torus so much that model-based \nhlos\ at 70--80$^{\circ}$ inclination could start matching the observed values. However, this requires significantly more computing time compared to the radio opacity calculations primarily because multiple scatterings become increasingly important as the equatorial plane is driven toward optical depths in the CT regime \citep[e.g.,][]{murphy+yaqoob-2009}. A direct fit to the radio measurements and a joint fit to both X-ray and radio data is therefore outside of the scope of this study and will be explored in future work.

For each model curve shown in Figure~\ref{fig:radioop}, we adjusted the outer radius of the underlying torus model to match the steep opacity decrease at 50\,\% of the peak around $+2$\,mas. The resulting outer radii are 0.25--0.5\,pc for the isotropic models and 0.5--1\,pc for the vertical gradient models. X-ray torus reprocessing models in general lack a physical scale (i.e., spectra only depend on the optical depth distribution), while current X-ray instruments can only partially resolve reprocessed emission extended on $\gtrsim\!100$\,pc scales in a limited number of nearby AGN \citep{fabbiano+2017-eso428-p1,ma+2020,jones+2021}. More self-consistent multiwavelength studies such as the one presented in this section may prove valuable for indirectly setting the spatial scale of the compact reprocessed emission region, informing the expectations for potential direct size measurements using X-ray interferometry in the future \citep{uttley+2019-wp}.

\subsection{Broader and Multiwavelength Context} 
\label{sec:discussion-mwl} 

Our multi-epoch X-ray modeling is based on a simple model for X-ray reprocessing in the circumnuclear material, which nevertheless represents a step forward compared to phenomenological models used in previous studies. The X-ray analysis alone leads us to conclude that NGC\,1052 lacks a torus containing dense, CT material.
Tori containing very little, if any, CT material may not be uncommon among LL\,AGN such as NGC\,1052 (\lamedd\,$\approx4\times10^{-4}$; \citealt{fernandezOntiveros+2019-ngc1052}). Also using the \borustwo\ model, \citet{diaz+2020-ngc3718} found that the reprocessing features in the broadband X-ray spectrum of NGC\,3718 (\lamedd$\simeq1\times10^{-5}$) are best accounted for by a torus with \lognhtorcmmt$<23.2$ and preferentially high covering factor. Using similar spectral models, equally low \nhtor\ constraints have recently been found for other LL\,AGN with high-quality \nustar\ observations, such as M\,81 (\lamedd$\simeq1\times10^{-5}$; \citealt{young+2018-m81}), NGC\,3998, and NGC\,4579 (\lamedd$\simeq1\times10^{-5}$ and \lamedd$\simeq1\times10^{-4}$, respectively; \citealt{younes+2019}).

Our comparison with VLBI measurements suggests that torus models with high-density material confined to the equatorial plane may be able to simultaneously account for both X-ray spectra and small-scale radio data. This is especially true if we allow for an entirely possible mild misalignment in inclination between the jet and torus axes. For simplicity, our analysis assumed equal inclination, but we note that X-ray modeling would not change noticeably with a change in torus inclination by $\pm10^{\circ}$. For better understanding of the physical interaction between the jet and the torus in NGC\,1052 and other LL\,AGN with appreciable jets, it will be critical to self-consistently compute observable signatures like in the recent simulations by \citet{fromm+2018}. Despite providing a good fit to a reliable set of X-ray data and a tantalizingly simple interpretation, our best-fit torus model does not represent a unique solution. In future work we will consider improved models featuring steeper vertical density gradients, making the torus more like a puffy disk. Disk-like models (with assumed low covering factors) were used to explain pc-scale absorption of receding jets in other LL\,AGN, such as NGC\,4261 \citep{haga+2015-ngc4261} and NGC\,1275 \citep{wajima+2020-ngc1275}.

Such considerations are important for understanding the dependence of the torus covering factor on intrinsic luminosity \citep{brightman+2015,balokovic-2017,marchesi+2019-cftor}, which is indirectly probed by the ratio of fraction of obscured AGN in the general population \citep[e.g.,][]{hasinger-2008,burlon+2011,vasudevan+2013}. The covering factor may depend more directly on \lamedd\ \citep{ricci+2017-cftor,she+2018}. The covering factor for dust, constrained from infrared observations, shows a qualitatively similar decrease toward high luminosities in some studies \citep{alonsoHerrero+2011}, while others find no significant trends \citep{stalevski+2016,garciaBernete+2019}. Results for LL\,AGN obtained from the X-ray and infrared bands are divergent and even less clear due to limited statistics \citep[e.g.,][]{kawamuro+2016-lowLum,gonzalezMartin+2017,ichikawa+2019}. Studying NGC\,1052 at the highest spatial resolution available in the infrared, \citet{fernandezOntiveros+2019-ngc1052} concluded that jets are the more likely source of broadband infrared emission at $\lesssim\!0.5$\,pc, hence finding no evidence for thermal emission from a dusty torus.

Theoretically, it has been suggested that the typical dense and dusty Seyfert torus cannot be adequately supported in the LL\,AGN regime \citep{elitzur+shlosman-2006,hoenig+beckert-2007,ramosAlmeida+ricci-2017}.
Established structural differences in the accretion flow at the smallest scales (i.e., hot accretion flow replacing a thin accretion disk) may be accompanied by differences on pc scales, which are typically associated with the torus, especially if the broad-line region is essentially a smooth continuation of the torus at its inner edge \citep[e.g.,][]{nandra-2006,shu+2011,gandhi+2015}. Since X-ray reprocessing is dominated by gas and not dust, constraints on the torus from the X-ray band (at energy resolution considered in our study) do not distinguish these two phenomenologically separated structures. Multiwavelength studies directly leveraging joint constraints from more than one band \citep[e.g.,][]{esparzaArredondo+2019-ic5063,lanz+2019,ogawa+2021,esparzaArredondo+2021}
are needed to elucidate the way forward in building more self-consistent physical models for the complex systems that hide behind the deceptively simple idea of the obscuring ``torus''.

\section{Summary} 
\label{sec:summary} 

In this paper we present modeling of multi-epoch broadband X-ray spectra of NGC\,1052 that is novel in at least three aspects: (i) we anchor it with hard X-ray data integrated over 4--8.5 years representing the long-term average spectrum, (ii) we separate variable components from constant components related to the average spectrum, and (iii) we use physically motivated models for X-ray reprocessing in the torus. The main result of our X-ray spectral analysis is a model that self-consistently describes four individual observations extending into the hard X-ray band, the average hard X-ray data, and the spectral variability observed in the 2--10\,keV band. Without prior assumption, the results suggest that the observed range of the line-of-sight column density closely matches the average column density of the torus, well below the CT threshold. The torus is found to have a covering factor of 80--100\,\%.

The straightforward interpretation that naturally emerges from our X-ray data analysis is consistent with other recent findings for the properties of LL\,AGN tori that suggest the lack of a significant covering factor of dense circumnuclear material with CT column density. However, this may be due to the limitations of the current X-ray reprocessing models, which require further development and input from multiwavelength data. In particular, we examined VLBI measurements of absorption on sub-pc scale, finding that a torus model with a steep density gradient roughly in the direction of jet axis may be able to account for both X-ray and radio observations. This provides a valuable direction for the development of improved AGN torus models that could provide a more physically self-consistent picture of the circumnuclear environment in NGC\,1052 and potentially in a wider class of AGN at the transition between LL\,AGN and Seyferts.

\acknowledgments 

The authors appreciate thoughtful suggestions from the anonymous referee, which helped to improve the clarity of the paper. We thank E. Ros and S. Sawada-Satoh for informative discussions on the radio observations of NGC\,1052, and S. Doeleman for unreserved support of the projects that led to this publication. M.\,B. and S.\,E.\,C. gratefully acknowledge support from the Black Hole Initiative at Harvard University, which is funded in part by the Gordon and Betty Moore Foundation (grant GBMF8273) and in part by the John Templeton Foundation. M.\,B. also acknowledges support from the \chandra\ award TM9-20004X issued by the \chandra\ X-ray Observatory Center and the YCAA Prize Postdoctoral Fellowship.

We have made use of data from the \nustar\ mission, a project led by the California Institute of Technology, managed by the Jet Propulsion Laboratory, and funded by the National Aeronautics and Space Administration. This research has made use of the \nustar\ Data Analysis Software (NuSTARDAS) jointly developed by the Space Science Data Center (SSDC; ASI, Italy) and the California Institute of Technology (USA). Part of this work is based on archival data, software or online services provided by the SSDC, and on observations obtained with \xmmnewton, an ESA science mission with instruments and contributions directly funded by ESA Member States and NASA. This research has made use of data and/or software provided by the High Energy Astrophysics Science Archive Research Center (HEASARC), which is a service of the Astrophysics Science Division at NASA/GSFC and the High Energy Astrophysics Division of the Smithsonian Astrophysical Observatory. The computations for this paper were partially conducted on the Smithsonian High-Performance Cluster (SI/HPC; \url{https://doi.org/10.25572/SIHPC}).

\facilities{\nustar, \xmmnewton, \suzaku, \bepposax, {\em Neil Gerhels Swift Observatory}, \integral, \rxte}

\software{\texttt{Astropy} \citep{astropy-2013,astropy-2018}, \texttt{Matplotlib} \citep{hunter+2007}, \xspec\ \citep{arnaud-1996}, \texttt{WebPlotDigitizer} (\url{https://apps.automeris.io/wpd/})}

\appendix

\section{Instrumental Cross-normalization Factors} 
\label{sec:appendix-cnfactors} 

In all models presented in \S\,\ref{sec:fitting-singleepoch}, including the well-fitting ones, we note that the cross-normalization factor (CNF) between the \nustar\ and \xmmnewton\ instruments is unexpectedly far from unity (0.85--0.87). A similar problem was identified by \citet{osorioClavijo+2020-ngc1052}, which they attributed to a mismatch in observed spectral slopes. This issue is persistent over multiple different data processing procedures. We find that except for the offset normalization, the \nustar\ and \xmmnewton\ spectra agree very well in the overlapping 3--12\,keV band covered by high-quality data. Adopting a well-fitting model over that band and allowing different $\Gamma$ for \nustar\ and \xmmnewton\ spectra converges to $\Gamma$ differing by about 0.1, without any significant effect on the CNF. Further testing is outside of the scope of this study, so we simply ignored the offset in the remainder of the spectral analysis presented in this paper.

In our multi-epoch analysis presented in \S\,\ref{sec:fitting-multiepoch}, as well as in multi-epoch models discussed in \S\,\ref{sec:discussion-torus}, we adopted a simple scheme for instrumental CNFs such that one hard X-ray instrument per epoch has unity CNF, while all others are determined from the data. In our default configuration, fixed, unity CFSs are \swiftbat\ (epoch 0), \nustar/FPMA (epochs 1 and 2), \suzaku/PIN (epoch 3), and \bepposax/PDS (epoch 4). Table~\ref{tab:cnfactors} lists the fitted CNFs for the spectral models listed in Table~\ref{tab:models}. Different choices for fixed versus fitted CNFs do not change any best-fit models. Fixing CNFs to values determined from cross-calibration \citep{madsen+2017-crossCal} results in small shifts in some spectral parameters (mainly $K$ and \nhlos) and generally higher $\chi^2$. We verified that none of the results presented in this paper would significantly change under an alternative CNF scheme and chose to mainly present the one that is the most flexible.

\begin{deluxetable*}{lcccc}
\tablecaption{Cross-normalization factors (CNF) for models listed in Table~\ref{tab:models} \label{tab:cnfactors}}
\tabletypesize{\small}
\tablehead{
  \colhead{CNF} &
  \colhead{Model~S} &
  \colhead{Model~A\arcmin} &
  \colhead{Model~B} &
  \colhead{Model~B$_3$}
}
\startdata
ISGRI & \nodata & $1.1\pm0.1$ & $1.1\pm0.1$ & $1.0\pm0.1$ \\
PCA & \nodata & $1.21\pm0.04$ & $1.23\pm0.04$ & $1.07\pm0.02$ \\
FPMB (2017) & $0.99\pm0.01$ & $1.00\pm0.01$ & $1.00\pm0.01$ & $0.99\pm0.01$ \\
PN & $0.862\pm0.008$ & $0.86\pm0.01$ & $0.86\pm0.01$ & $0.863\pm0.008$ \\
MOS1+2 & $0.865\pm0.008$ & $0.86\pm0.01$ & $0.86\pm0.01$ & $0.862\pm0.009$ \\
FPMB (2013) & \nodata & $1.05\pm0.03$ & $1.05\pm0.03$ & $1.05\pm0.02$ \\
XIS1 & \nodata & $1.01\pm0.03$ & $1.01\pm0.03$ & $0.98\pm0.02$ \\
XIS0+3 & \nodata & $1.03\pm0.03$ & $1.02\pm0.03$ & $0.99\pm0.02$ \\
MECS2+3 & \nodata & $1.3\pm0.2$ & $1.4\pm0.2$ & \nodata \\
LECS & \nodata & $1.1\pm0.2$ & $1.2\pm0.2$ & \nodata \\
\enddata
\end{deluxetable*}

\bibliographystyle{aasjournal} 
\bibliography{aaa} 

\begin{thebibliography}{}
\expandafter\ifx\csname natexlab\endcsname\relax\def\natexlab#1{#1}\fi
\providecommand{\url}[1]{\href{#1}{#1}}

\bibitem[{{Alonso-Herrero} {et~al.}(2011){Alonso-Herrero}, {Ramos Almeida},
  {Mason}, {Asensio Ramos}, {Roche}, {Levenson}, {Elitzur}, {Packham},
  {Rodr{\'{\i}}guez Espinosa}, {Young}, {D{\'{\i}}az-Santos}, \&
  {P{\'e}rez-Garc{\'{\i}}a}}]{alonsoHerrero+2011}
{Alonso-Herrero}, A., {Ramos Almeida}, C., {Mason}, R., {et~al.} 2011, \apj,
  736, 82

\bibitem[{{Antonucci}(1993)}]{antonucci-1993}
{Antonucci}, R. 1993, \araa, 31, 473

\bibitem[{{Arnaud}(1996)}]{arnaud-1996}
{Arnaud}, K.~A. 1996, in Astronomical Society of the Pacific Conference Series,
  Vol. 101, Astronomical Data Analysis Software and Systems V, ed. G.~H.
  {Jacoby} \& J.~{Barnes}, 17

\bibitem[{{Astropy Collaboration} {et~al.}(2013){Astropy Collaboration},
  {Robitaille}, {Tollerud}, {Greenfield}, {Droettboom}, {Bray}, {Aldcroft},
  {Davis}, {Ginsburg}, {Price-Whelan}, {Kerzendorf}, {Conley}, {Crighton},
  {Barbary}, {Muna}, {Ferguson}, {Grollier}, {Parikh}, {Nair}, {Unther},
  {Deil}, {Woillez}, {Conseil}, {Kramer}, {Turner}, {Singer}, {Fox}, {Weaver},
  {Zabalza}, {Edwards}, {Azalee Bostroem}, {Burke}, {Casey}, {Crawford},
  {Dencheva}, {Ely}, {Jenness}, {Labrie}, {Lim}, {Pierfederici}, {Pontzen},
  {Ptak}, {Refsdal}, {Servillat}, \& {Streicher}}]{astropy-2013}
{Astropy Collaboration}, {Robitaille}, T.~P., {Tollerud}, E.~J., {et~al.} 2013,
  \aap, 558, A33

\bibitem[{{Astropy Collaboration} {et~al.}(2018){Astropy Collaboration},
  {Price-Whelan}, {Sip{\H o}cz}, {G{\"u}nther}, {Lim}, {Crawford}, {Conseil},
  {Shupe}, {Craig}, {Dencheva}, {Ginsburg}, {VanderPlas}, {Bradley},
  {P{\'e}rez-Su{\'a}rez}, {de Val-Borro}, {Aldcroft}, {Cruz}, {Robitaille},
  {Tollerud}, {Ardelean}, {Babej}, {Bach}, {Bachetti}, {Bakanov}, {Bamford},
  {Barentsen}, {Barmby}, {Baumbach}, {Berry}, {Biscani}, {Boquien}, {Bostroem},
  {Bouma}, {Brammer}, {Bray}, {Breytenbach}, {Buddelmeijer}, {Burke},
  {Calderone}, {Cano Rodr{\'{\i}}guez}, {Cara}, {Cardoso}, {Cheedella},
  {Copin}, {Corrales}, {Crichton}, {D'Avella}, {Deil}, {Depagne}, {Dietrich},
  {Donath}, {Droettboom}, {Earl}, {Erben}, {Fabbro}, {Ferreira}, {Finethy},
  {Fox}, {Garrison}, {Gibbons}, {Goldstein}, {Gommers}, {Greco}, {Greenfield},
  {Groener}, {Grollier}, {Hagen}, {Hirst}, {Homeier}, {Horton}, {Hosseinzadeh},
  {Hu}, {Hunkeler}, {Ivezi{\'c}}, {Jain}, {Jenness}, {Kanarek}, {Kendrew},
  {Kern}, {Kerzendorf}, {Khvalko}, {King}, {Kirkby}, {Kulkarni}, {Kumar},
  {Lee}, {Lenz}, {Littlefair}, {Ma}, {Macleod}, {Mastropietro}, {McCully},
  {Montagnac}, {Morris}, {Mueller}, {Mumford}, {Muna}, {Murphy}, {Nelson},
  {Nguyen}, {Ninan}, {N{\"o}the}, {Ogaz}, {Oh}, {Parejko}, {Parley}, {Pascual},
  {Patil}, {Patil}, {Plunkett}, {Prochaska}, {Rastogi}, {Reddy Janga},
  {Sabater}, {Sakurikar}, {Seifert}, {Sherbert}, {Sherwood-Taylor}, {Shih},
  {Sick}, {Silbiger}, {Singanamalla}, {Singer}, {Sladen}, {Sooley},
  {Sornarajah}, {Streicher}, {Teuben}, {Thomas}, {Tremblay}, {Turner},
  {Terr{\'o}n}, {van Kerkwijk}, {de la Vega}, {Watkins}, {Weaver}, {Whitmore},
  {Woillez}, {Zabalza}, \& {Astropy Contributors}}]{astropy-2018}
{Astropy Collaboration}, {Price-Whelan}, A.~M., {Sip{\H o}cz}, B.~M., {et~al.}
  2018, \aj, 156, 123

\bibitem[{{Baczko} {et~al.}(2019){Baczko}, {Schulz}, {Kadler}, {Ros},
  {Perucho}, {Fromm}, \& {Wilms}}]{baczko+2019-ngc1052}
{Baczko}, A.~K., {Schulz}, R., {Kadler}, M., {et~al.} 2019, \aap, 623, A27

\bibitem[{{Baczko} {et~al.}(2016){Baczko}, {Schulz}, {Kadler}, {Ros},
  {Perucho}, {Krichbaum}, {B{\"o}ck}, {Bremer}, {Grossberger}, {Lindqvist},
  {Lobanov}, {Mannheim}, {Mart{\'\i}-Vidal}, {M{\"u}ller}, {Wilms}, \&
  {Zensus}}]{baczko+2016-ngc1052}
---. 2016, \aap, 593, A47

\bibitem[{{Balokovi\'{c}}(2017)}]{balokovic-2017}
{Balokovi\'{c}}, M. 2017, PhD thesis, California Institute of Technology,
  doi:10.7907/Z9WM1BG8

\bibitem[{{Balokovi\'{c}} {et~al.}(2019){Balokovi\'{c}}, {Garc{\'\i}a}, \&
  {Cabral}}]{balokovic+2019-rnaas}
{Balokovi\'{c}}, M., {Garc{\'\i}a}, J.~A., \& {Cabral}, S.~E. 2019, Research
  Notes of the American Astronomical Society, 3, 173

\bibitem[{{Balokovi\'{c}} {et~al.}(2018){Balokovi\'{c}}, {Brightman},
  {Harrison}, {Comastri}, {Ricci}, {Buchner}, {Gandhi}, {Farrah}, \&
  {Stern}}]{balokovic+2018}
{Balokovi\'{c}}, M., {Brightman}, M., {Harrison}, F.~A., {et~al.} 2018, \apj,
  854, 42

\bibitem[{{Balokovi{\'c}} {et~al.}(2020){Balokovi{\'c}}, {Harrison},
  {Madejski}, {Comastri}, {Ricci}, {Annuar}, {Ballantyne}, {Boorman}, {Brandt},
  {Brightman}, {Gandhi}, {Kamraj}, {Koss}, {Marchesi}, {Marinucci}, {Masini},
  {Matt}, {Stern}, \& {Urry}}]{balokovic+2020}
{Balokovi{\'c}}, M., {Harrison}, F.~A., {Madejski}, G., {et~al.} 2020, \apj,
  905, 41

\bibitem[{{Barth} {et~al.}(1999){Barth}, {Filippenko}, \&
  {Moran}}]{barth+1999-ngc1052}
{Barth}, A.~J., {Filippenko}, A.~V., \& {Moran}, E.~C. 1999, \apjl, 515, L61

\bibitem[{{Barthelmy} {et~al.}(2005){Barthelmy}, {Barbier}, {Cummings},
  {Fenimore}, {Gehrels}, {Hullinger}, {Krimm}, {Markwardt}, {Palmer},
  {Parsons}, {Sato}, {Suzuki}, {Takahashi}, {Tashiro}, \&
  {Tueller}}]{barthelmy+2005-bat}
{Barthelmy}, S.~D., {Barbier}, L.~M., {Cummings}, J.~R., {et~al.} 2005, SSRv,
  120, 143

\bibitem[{{Boella} {et~al.}(1997{\natexlab{a}}){Boella}, {Butler}, {Perola},
  {Piro}, {Scarsi}, \& {Bleeker}}]{boella+1997-bepposax}
{Boella}, G., {Butler}, R.~C., {Perola}, G.~C., {et~al.} 1997{\natexlab{a}},
  \aaps, 122, 299

\bibitem[{{Boella} {et~al.}(1997{\natexlab{b}}){Boella}, {Chiappetti}, {Conti},
  {Cusumano}, {del Sordo}, {La Rosa}, {Maccarone}, {Mineo}, {Molendi}, {Re},
  {Sacco}, \& {Tripiciano}}]{boella+1997-bepposaxmecs}
{Boella}, G., {Chiappetti}, L., {Conti}, G., {et~al.} 1997{\natexlab{b}},
  \aaps, 122, 327

\bibitem[{{Bradt} {et~al.}(1993){Bradt}, {Rothschild}, \&
  {Swank}}]{bradt+1993-rxte}
{Bradt}, H.~V., {Rothschild}, R.~E., \& {Swank}, J.~H. 1993, \aaps, 97, 355

\bibitem[{{Brenneman} {et~al.}(2009){Brenneman}, {Weaver}, {Kadler}, {Tueller},
  {Marscher}, {Ros}, {Zensus}, {Kovalev}, {Aller}, {Aller}, {Irwin}, {Kerp}, \&
  {Kaufmann}}]{brenneman+2009-ngc1052}
{Brenneman}, L.~W., {Weaver}, K.~A., {Kadler}, M., {et~al.} 2009, \apj, 698,
  528

\bibitem[{{Brightman} {et~al.}(2015){Brightman}, {Balokovi{\'c}}, {Stern},
  {Ar{\'e}valo}, {Ballantyne}, {Bauer}, {Boggs}, {Craig}, {Christensen},
  {Comastri}, {Fuerst}, {Gandhi}, {Hailey}, {Harrison}, {Hickox}, {Koss},
  {LaMassa}, {Puccetti}, {Rivers}, {Vasudevan}, {Walton}, \&
  {Zhang}}]{brightman+2015}
{Brightman}, M., {Balokovi{\'c}}, M., {Stern}, D., {et~al.} 2015, \apj, 805, 41

\bibitem[{{Buchner} {et~al.}(2019){Buchner}, {Brightman}, {Nandra}, {Nikutta},
  \& {Bauer}}]{buchner+2019}
{Buchner}, J., {Brightman}, M., {Nandra}, K., {Nikutta}, R., \& {Bauer}, F.~E.
  2019, \aap, 629, A16

\bibitem[{{Burlon} {et~al.}(2011){Burlon}, {Ajello}, {Greiner}, {Comastri},
  {Merloni}, \& {Gehrels}}]{burlon+2011}
{Burlon}, D., {Ajello}, M., {Greiner}, J., {et~al.} 2011, \apj, 728, 58

\bibitem[{{Cabral}(2020)}]{cabral-2020}
{Cabral}, S.~E. 2020, Master's thesis, University of Massachusetts Boston,
  doi:https://scholarworks.umb.edu/masters\_theses/630

\bibitem[{{Claussen} {et~al.}(1998){Claussen}, {Diamond}, {Braatz}, {Wilson},
  \& {Henkel}}]{claussen+1998-ngc1052}
{Claussen}, M.~J., {Diamond}, P.~J., {Braatz}, J.~A., {Wilson}, A.~S., \&
  {Henkel}, C. 1998, \apjl, 500, L129

\bibitem[{{Connolly} {et~al.}(2016){Connolly}, {McHardy}, {Skipper}, \&
  {Emmanoulopoulos}}]{connolly+2016}
{Connolly}, S.~D., {McHardy}, I.~M., {Skipper}, C.~J., \& {Emmanoulopoulos}, D.
  2016, \mnras, 459, 3963

\bibitem[{{Constantin} {et~al.}(2009){Constantin}, {Green}, {Aldcroft}, {Kim},
  {Haggard}, {Barkhouse}, \& {Anderson}}]{constantin+2009}
{Constantin}, A., {Green}, P., {Aldcroft}, T., {et~al.} 2009, \apj, 705, 1336

\bibitem[{{Dauser} {et~al.}(2014){Dauser}, {Garcia}, {Parker}, {Fabian}, \&
  {Wilms}}]{dauser+2014-relxill}
{Dauser}, T., {Garcia}, J., {Parker}, M.~L., {Fabian}, A.~C., \& {Wilms}, J.
  2014, \mnras, 444, L100

\bibitem[{{Denicol{\'o}} {et~al.}(2005){Denicol{\'o}}, {Terlevich},
  {Terlevich}, {Forbes}, {Terlevich}, \& {Carrasco}}]{denicolo+2005}
{Denicol{\'o}}, G., {Terlevich}, R., {Terlevich}, E., {et~al.} 2005, \mnras,
  356, 1440

\bibitem[{{Diaz} {et~al.}(2020){Diaz}, {Ar{\'e}valo},
  {Hern{\'a}ndez-Garc{\'\i}a}, {Bassani}, {Malizia},
  {Gonz{\'a}lez-Mart{\'\i}n}, {Ricci}, {Matt}, {Stern}, {May}, {Zezas}, \&
  {Bauer}}]{diaz+2020-ngc3718}
{Diaz}, Y., {Ar{\'e}valo}, P., {Hern{\'a}ndez-Garc{\'\i}a}, L., {et~al.} 2020,
  \mnras, 496, 5399

\bibitem[{{Elitzur} \& {Shlosman}(2006)}]{elitzur+shlosman-2006}
{Elitzur}, M., \& {Shlosman}, I. 2006, \apjl, 648, L101

\bibitem[{{Esparza Arredondo} {et~al.}(2021){Esparza Arredondo}, {Gonz{\'a}lez
  Mart{\'\i}n}, {Dultzin}, {Masegosa}, {Ramos Almeida}, {Garc{\'\i}a Bernete},
  {Fritz}, \& {Osorio Clavijo}}]{esparzaArredondo+2021}
{Esparza Arredondo}, D., {Gonz{\'a}lez Mart{\'\i}n}, O., {Dultzin}, D.,
  {et~al.} 2021, arXiv e-prints, arXiv:2104.11263

\bibitem[{{Esparza-Arredondo} {et~al.}(2019){Esparza-Arredondo},
  {Gonz{\'a}lez-Mart{\'\i}n}, {Dultzin}, {Ramos Almeida}, {Fritz}, {Masegosa},
  {Pasetto}, {Mart{\'\i}nez-Paredes}, {Osorio-Clavijo}, \&
  {Victoria-Ceballos}}]{esparzaArredondo+2019-ic5063}
{Esparza-Arredondo}, D., {Gonz{\'a}lez-Mart{\'\i}n}, O., {Dultzin}, D.,
  {et~al.} 2019, \apj, 886, 125

\bibitem[{{Fabbiano} {et~al.}(2017){Fabbiano}, {Elvis}, {Paggi}, {Karovska},
  {Maksym}, {Raymond}, {Risaliti}, \& {Wang}}]{fabbiano+2017-eso428-p1}
{Fabbiano}, G., {Elvis}, M., {Paggi}, A., {et~al.} 2017, \apjl, 842, L4

\bibitem[{{Falocco} {et~al.}(2020){Falocco}, {Larsson}, \&
  {Nandi}}]{falocco+2020-ngc1052}
{Falocco}, S., {Larsson}, J., \& {Nandi}, S. 2020, \aap, 638, A67

\bibitem[{{Fern{\'a}ndez-Ontiveros} {et~al.}(2019){Fern{\'a}ndez-Ontiveros},
  {L{\'o}pez-Gonzaga}, {Prieto}, {Acosta-Pulido}, {Lopez-Rodriguez}, {Asmus},
  \& {Tristram}}]{fernandezOntiveros+2019-ngc1052}
{Fern{\'a}ndez-Ontiveros}, J.~A., {L{\'o}pez-Gonzaga}, N., {Prieto}, M.~A.,
  {et~al.} 2019, \mnras, 485, 5377

\bibitem[{{Fromm} {et~al.}(2018){Fromm}, {Perucho}, {Porth}, {Younsi}, {Ros},
  {Mizuno}, {Zensus}, \& {Rezzolla}}]{fromm+2018}
{Fromm}, C.~M., {Perucho}, M., {Porth}, O., {et~al.} 2018, \aap, 609, A80

\bibitem[{{Fromm} {et~al.}(2019){Fromm}, {Younsi}, {Baczko}, {Mizuno}, {Porth},
  {Perucho}, {Olivares}, {Nathanail}, {Angelakis}, {Ros}, {Zensus}, \&
  {Rezzolla}}]{fromm+2019-ngc1052}
{Fromm}, C.~M., {Younsi}, Z., {Baczko}, A., {et~al.} 2019, \aap, 629, A4

\bibitem[{{Frontera} {et~al.}(1997){Frontera}, {Costa}, {dal Fiume}, {Feroci},
  {Nicastro}, {Orlandini}, {Palazzi}, \&
  {Zavattini}}]{frontera+1997-bepposaxpds}
{Frontera}, F., {Costa}, E., {dal Fiume}, D., {et~al.} 1997, \aaps, 122, 357

\bibitem[{{Gabriel} {et~al.}(2004){Gabriel}, {Denby}, {Fyfe}, {Hoar}, {Ibarra},
  {Ojero}, {Osborne}, {Saxton}, {Lammers}, \& {Vacanti}}]{gabriel+2004-xmmsas}
{Gabriel}, C., {Denby}, M., {Fyfe}, D.~J., {et~al.} 2004, in Astronomical
  Society of the Pacific Conference Series, Vol. 314, Astronomical Data
  Analysis Software and Systems (ADASS) XIII, ed. F.~{Ochsenbein}, M.~G.
  {Allen}, \& D.~{Egret}, 759

\bibitem[{{Gandhi} {et~al.}(2015){Gandhi}, {H{\"o}nig}, \&
  {Kishimoto}}]{gandhi+2015}
{Gandhi}, P., {H{\"o}nig}, S.~F., \& {Kishimoto}, M. 2015, \apj, 812, 113

\bibitem[{{Garc{\'\i}a} {et~al.}(2014){Garc{\'\i}a}, {Dauser}, {Lohfink},
  {Kallman}, {Steiner}, {McClintock}, {Brenneman}, {Wilms}, {Eikmann},
  {Reynolds}, \& {Tombesi}}]{garcia+2014-relxill}
{Garc{\'\i}a}, J., {Dauser}, T., {Lohfink}, A., {et~al.} 2014, \apj, 782, 76

\bibitem[{{Garc{\'\i}a-Bernete} {et~al.}(2019){Garc{\'\i}a-Bernete}, {Ramos
  Almeida}, {Alonso-Herrero}, {Ward}, {Acosta-Pulido}, {Pereira-Santaella},
  {Hern{\'a}n-Caballero}, {Asensio Ramos}, {Gonz{\'a}lez-Mart{\'\i}n},
  {Levenson}, {Mateos}, {Carrera}, {Ricci}, {Roche}, {Marquez}, {Packham},
  {Masegosa}, \& {Fuller}}]{garciaBernete+2019}
{Garc{\'\i}a-Bernete}, I., {Ramos Almeida}, C., {Alonso-Herrero}, A., {et~al.}
  2019, \mnras, 486, 4917

\bibitem[{{Gehrels} {et~al.}(2004){Gehrels}, {Chincarini}, {Giommi}, {Mason},
  {Nousek}, {Wells}, {White}, {Barthelmy}, {Burrows}, {Cominsky}, {Hurley},
  {Marshall}, {M{\'e}sz{\'a}ros}, {Roming}, {Angelini}, {Barbier}, {Belloni},
  {Campana}, {Caraveo}, {Chester}, {Citterio}, {Cline}, {Cropper}, {Cummings},
  {Dean}, {Feigelson}, {Fenimore}, {Frail}, {Fruchter}, {Garmire}, {Gendreau},
  {Ghisellini}, {Greiner}, {Hill}, {Hunsberger}, {Krimm}, {Kulkarni}, {Kumar},
  {Lebrun}, {Lloyd-Ronning}, {Markwardt}, {Mattson}, {Mushotzky}, {Norris},
  {Osborne}, {Paczynski}, {Palmer}, {Park}, {Parsons}, {Paul}, {Rees},
  {Reynolds}, {Rhoads}, {Sasseen}, {Schaefer}, {Short}, {Smale}, {Smith},
  {Stella}, {Tagliaferri}, {Takahashi}, {Tashiro}, {Townsley}, {Tueller},
  {Turner}, {Vietri}, {Voges}, {Ward}, {Willingale}, {Zerbi}, \&
  {Zhang}}]{gehrels+2004}
{Gehrels}, N., {Chincarini}, G., {Giommi}, P., {et~al.} 2004, \apj, 611, 1005

\bibitem[{{Giann{\'\i}} {et~al.}(2011){Giann{\'\i}}, {de Rosa}, {Bassani},
  {Bazzano}, {Dean}, \& {Ubertini}}]{gianni+2011}
{Giann{\'\i}}, S., {de Rosa}, A., {Bassani}, L., {et~al.} 2011, \mnras, 411,
  2137

\bibitem[{{Gonz{\'a}lez-Mart{\'\i}n} {et~al.}(2009){Gonz{\'a}lez-Mart{\'\i}n},
  {Masegosa}, {M{\'a}rquez}, {Guainazzi}, \&
  {Jim{\'e}nez-Bail{\'o}n}}]{gonzalezMartin+2009}
{Gonz{\'a}lez-Mart{\'\i}n}, O., {Masegosa}, J., {M{\'a}rquez}, I., {Guainazzi},
  M., \& {Jim{\'e}nez-Bail{\'o}n}, E. 2009, \aap, 506, 1107

\bibitem[{{Gonz{\'a}lez-Mart{\'\i}n} {et~al.}(2017){Gonz{\'a}lez-Mart{\'\i}n},
  {Masegosa}, {Hern{\'a}n-Caballero}, {M{\'a}rquez}, {Ramos Almeida},
  {Alonso-Herrero}, {Aretxaga}, {Rodr{\'\i}guez-Espinosa}, {Acosta-Pulido},
  {Hern{\'a}ndez-Garc{\'\i}a}, {Esparza-Arredondo}, {Mart{\'\i}nez-Paredes},
  {Bonfini}, {Pasetto}, \& {Dultzin}}]{gonzalezMartin+2017}
{Gonz{\'a}lez-Mart{\'\i}n}, O., {Masegosa}, J., {Hern{\'a}n-Caballero}, A.,
  {et~al.} 2017, \apj, 841, 37

\bibitem[{{Guainazzi} {et~al.}(2000){Guainazzi}, {Oosterbroek}, {Antonelli}, \&
  {Matt}}]{guainazzi+2000-ngc1052}
{Guainazzi}, M., {Oosterbroek}, T., {Antonelli}, L.~A., \& {Matt}, G. 2000,
  \aap, 364, L80

\bibitem[{{Guainazzi} {et~al.}(2016){Guainazzi}, {Risaliti}, {Awaki},
  {Arevalo}, {Bauer}, {Bianchi}, {Boggs}, {Brandt}, {Brightman}, {Christensen},
  {Craig}, {Forster}, {Hailey}, {Harrison}, {Koss}, {Longinotti}, {Markwardt},
  {Marinucci}, {Matt}, {Reynolds}, {Ricci}, {Stern}, {Svoboda}, {Walton}, \&
  {Zhang}}]{guainazzi+2016-mrk3}
{Guainazzi}, M., {Risaliti}, G., {Awaki}, H., {et~al.} 2016, \mnras, 460, 1954

\bibitem[{{Gupta} {et~al.}(2021){Gupta}, {Ricci}, {Tortosa}, {Ueda},
  {Kawamuro}, {Koss}, {Trakhtenbrot}, {Oh}, {Bauer}, {Ricci}, {Privon},
  {Zappacosta}, {Stern}, {Kakkad}, {Piconcelli}, {Veilleux}, {Mushotzky},
  {Caglar}, {Ichikawa}, {Elagali}, {Powell}, {Urry}, \&
  {Harrison}}]{gupta+2021}
{Gupta}, K.~K., {Ricci}, C., {Tortosa}, A., {et~al.} 2021, \mnras, 504, 428

\bibitem[{{Haga} {et~al.}(2015){Haga}, {Doi}, {Murata}, {Sudou}, {Kameno}, \&
  {Hada}}]{haga+2015-ngc4261}
{Haga}, T., {Doi}, A., {Murata}, Y., {et~al.} 2015, \apj, 807, 15

\bibitem[{{Harrison} {et~al.}(2013){Harrison}, {Craig}, {Christensen},
  {Hailey}, {Zhang}, {Boggs}, {Stern}, {Cook}, {Forster}, {Giommi},
  {Grefenstette}, {Kim}, {Kitaguchi}, {Koglin}, {Madsen}, {Mao}, {Miyasaka},
  {Mori}, {Perri}, {Pivovaroff}, {Puccetti}, {Rana}, {Westergaard}, {Willis},
  {Zoglauer}, {An}, {Bachetti}, {Barri{\`e}re}, {Bellm}, {Bhalerao},
  {Brejnholt}, {Fuerst}, {Liebe}, {Markwardt}, {Nynka}, {Vogel}, {Walton},
  {Wik}, {Alexander}, {Cominsky}, {Hornschemeier}, {Hornstrup}, {Kaspi},
  {Madejski}, {Matt}, {Molendi}, {Smith}, {Tomsick}, {Ajello}, {Ballantyne},
  {Balokovi{\'c}}, {Barret}, {Bauer}, {Blandford}, {Brandt}, {Brenneman},
  {Chiang}, {Chakrabarty}, {Chenevez}, {Comastri}, {Dufour}, {Elvis}, {Fabian},
  {Farrah}, {Fryer}, {Gotthelf}, {Grindlay}, {Helfand}, {Krivonos}, {Meier},
  {Miller}, {Natalucci}, {Ogle}, {Ofek}, {Ptak}, {Reynolds}, {Rigby},
  {Tagliaferri}, {Thorsett}, {Treister}, \& {Urry}}]{harrison+2013}
{Harrison}, F.~A., {Craig}, W.~W., {Christensen}, F.~E., {et~al.} 2013, \apj,
  770, 103

\bibitem[{{Hasinger}(2008)}]{hasinger-2008}
{Hasinger}, G. 2008, \aap, 490, 905

\bibitem[{{Hern{\'a}ndez-Garc{\'\i}a}
  {et~al.}(2013){Hern{\'a}ndez-Garc{\'\i}a}, {Gonz{\'a}lez-Mart{\'\i}n},
  {M{\'a}rquez}, \& {Masegosa}}]{hernandezGarcia+2013-liners}
{Hern{\'a}ndez-Garc{\'\i}a}, L., {Gonz{\'a}lez-Mart{\'\i}n}, O., {M{\'a}rquez},
  I., \& {Masegosa}, J. 2013, \aap, 556, A47

\bibitem[{{HI4PI Collaboration} {et~al.}(2016){HI4PI Collaboration}, {Ben
  Bekhti}, {Fl{\"o}er}, {Keller}, {Kerp}, {Lenz}, {Winkel}, {Bailin},
  {Calabretta}, {Dedes}, {Ford}, {Gibson}, {Haud}, {Janowiecki}, {Kalberla},
  {Lockman}, {McClure-Griffiths}, {Murphy}, {Nakanishi}, {Pisano}, \&
  {Staveley-Smith}}]{hi4pi-2016-nhgal}
{HI4PI Collaboration}, {Ben Bekhti}, N., {Fl{\"o}er}, L., {et~al.} 2016, \aap,
  594, A116

\bibitem[{{Ho}(2008)}]{ho-2008}
{Ho}, L.~C. 2008, \araa, 46, 475

\bibitem[{{H{\"o}nig}(2019)}]{hoenig-2019}
{H{\"o}nig}, S.~F. 2019, \apj, 884, 171

\bibitem[{{H{\"o}nig} \& {Beckert}(2007)}]{hoenig+beckert-2007}
{H{\"o}nig}, S.~F., \& {Beckert}, T. 2007, \mnras, 380, 1172

\bibitem[{Hunter(2007)}]{hunter+2007}
Hunter, J.~D. 2007, Computing In Science \& Engineering, 9, 90

\bibitem[{{Ichikawa} {et~al.}(2019){Ichikawa}, {Ricci}, {Ueda}, {Bauer},
  {Kawamuro}, {Koss}, {Oh}, {Rosario}, {Shimizu}, {Stalevski}, {Fuller},
  {Packham}, \& {Trakhtenbrot}}]{ichikawa+2019}
{Ichikawa}, K., {Ricci}, C., {Ueda}, Y., {et~al.} 2019, \apj, 870, 31

\bibitem[{{Impellizzeri} {et~al.}(2008){Impellizzeri}, {Roy}, \&
  {Henkel}}]{impellizzeri+2008-ngc1052}
{Impellizzeri}, V., {Roy}, A.~L., \& {Henkel}, C. 2008, in The role of VLBI in
  the Golden Age for Radio Astronomy, Vol.~9, 33

\bibitem[{{Jahoda} {et~al.}(2006){Jahoda}, {Markwardt}, {Radeva}, {Rots},
  {Stark}, {Swank}, {Strohmayer}, \& {Zhang}}]{jahoda+2006-rxtepca}
{Jahoda}, K., {Markwardt}, C.~B., {Radeva}, Y., {et~al.} 2006, \apjs, 163, 401

\bibitem[{{Jansen} {et~al.}(2001){Jansen}, {Lumb}, {Altieri}, {Clavel}, {Ehle},
  {Erd}, {Gabriel}, {Guainazzi}, {Gondoin}, \& {Much}}]{jansen+2001-xmm}
{Jansen}, F., {Lumb}, D., {Altieri}, B., {et~al.} 2001, \aap, 365, L1

\bibitem[{{Jones} {et~al.}(2021){Jones}, {Parker}, {Fabbiano}, {Elvis},
  {Maksym}, {Paggi}, {Ma}, {Karovska}, {Siemiginowska}, \& {Wang}}]{jones+2021}
{Jones}, M.~L., {Parker}, K., {Fabbiano}, G., {et~al.} 2021, \apj, 910, 19

\bibitem[{{Kadler} {et~al.}(2004{\natexlab{a}}){Kadler}, {Kerp}, {Ros},
  {Falcke}, {Pogge}, \& {Zensus}}]{kadler+2004a-ngc1052-mwl}
{Kadler}, M., {Kerp}, J., {Ros}, E., {et~al.} 2004{\natexlab{a}}, \aap, 420,
  467

\bibitem[{{Kadler} {et~al.}(2004{\natexlab{b}}){Kadler}, {Ros}, {Lobanov},
  {Falcke}, \& {Zensus}}]{kadler+2004b-ngc1052-vlbi}
{Kadler}, M., {Ros}, E., {Lobanov}, A.~P., {Falcke}, H., \& {Zensus}, J.~A.
  2004{\natexlab{b}}, \aap, 426, 481

\bibitem[{{Kameno} {et~al.}(2001){Kameno}, {Sawada-Satoh}, {Inoue}, {Shen}, \&
  {Wajima}}]{kameno+2001-ngc1052}
{Kameno}, S., {Sawada-Satoh}, S., {Inoue}, M., {Shen}, Z.-Q., \& {Wajima}, K.
  2001, \pasj, 53, 169

\bibitem[{{Kawamuro} {et~al.}(2016){Kawamuro}, {Ueda}, {Tazaki}, {Terashima},
  \& {Mushotzky}}]{kawamuro+2016-lowLum}
{Kawamuro}, T., {Ueda}, Y., {Tazaki}, F., {Terashima}, Y., \& {Mushotzky}, R.
  2016, \apj, 831, 37

\bibitem[{{Koyama} {et~al.}(2007){Koyama}, {Tsunemi}, {Dotani}, {Bautz},
  {Hayashida}, {Tsuru}, {Matsumoto}, {Ogawara}, {Ricker}, {Doty}, {Kissel},
  {Foster}, {Nakajima}, {Yamaguchi}, {Mori}, {Sakano}, {Hamaguchi},
  {Nishiuchi}, {Miyata}, {Torii}, {Namiki}, {Katsuda}, {Matsuura}, {Miyauchi},
  {Anabuki}, {Tawa}, {Ozaki}, {Murakami}, {Maeda}, {Ichikawa}, {Prigozhin},
  {Boughan}, {Lamarr}, {Miller}, {Burke}, {Gregory}, {Pillsbury}, {Bamba},
  {Hiraga}, {Senda}, {Katayama}, {Kitamoto}, {Tsujimoto}, {Kohmura}, {Tsuboi},
  \& {Awaki}}]{koyama+2007-suzakuxis}
{Koyama}, K., {Tsunemi}, H., {Dotani}, T., {et~al.} 2007, \pasj, 59, 23

\bibitem[{{Laha} {et~al.}(2020){Laha}, {Markowitz}, {Krumpe}, {Nikutta},
  {Rothschild}, \& {Saha}}]{laha+2020}
{Laha}, S., {Markowitz}, A.~G., {Krumpe}, M., {et~al.} 2020, \apj, 897, 66

\bibitem[{{Lanz} {et~al.}(2019){Lanz}, {Hickox}, {Balokovi{\'c}}, {Shimizu},
  {Ricci}, {Goulding}, {Ballantyne}, {Bauer}, {Chen}, {del Moro}, {Farrah},
  {Michael}, {Koss}, {LaMassa}, {Masini}, \& {Zappacosta}}]{lanz+2019}
{Lanz}, L., {Hickox}, R.~C., {Balokovi{\'c}}, M., {et~al.} 2019, \apj, 870, 26

\bibitem[{{Lebrun} {et~al.}(2003){Lebrun}, {Leray}, {Lavocat}, {Cr{\'e}tolle},
  {Arqu{\`e}s}, {Blondel}, {Bonnin}, {Bou{\`e}re}, {Cara}, {Chaleil}, {Daly},
  {Desages}, {Dzitko}, {Horeau}, {Laurent}, {Limousin}, {Mathy}, {Mauguen},
  {Meignier}, {Molini{\'e}}, {Poindron}, {Rouger}, {Sauvageon}, \&
  {Tourrette}}]{lebrun+2003-isgri}
{Lebrun}, F., {Leray}, J.~P., {Lavocat}, P., {et~al.} 2003, \aap, 411, L141

\bibitem[{{Liu} \& {Li}(2014)}]{liu+li-2014}
{Liu}, Y., \& {Li}, X. 2014, \apj, 787, 52

\bibitem[{{Ma} {et~al.}(2020){Ma}, {Elvis}, {Fabbiano}, {Balokovi{\'c}},
  {Maksym}, {Jones}, \& {Risaliti}}]{ma+2020}
{Ma}, J., {Elvis}, M., {Fabbiano}, G., {et~al.} 2020, \apj, 900, 164

\bibitem[{{Madsen} {et~al.}(2017){Madsen}, {Beardmore}, {Forster}, {Guainazzi},
  {Marshall}, {Miller}, {Page}, \& {Stuhlinger}}]{madsen+2017-crossCal}
{Madsen}, K.~K., {Beardmore}, A.~P., {Forster}, K., {et~al.} 2017, \aj, 153, 2

\bibitem[{{Magdziarz} \& {Zdziarski}(1995)}]{magdziarz+zdziarski-1995}
{Magdziarz}, P., \& {Zdziarski}, A.~A. 1995, \mnras, 273, 837

\bibitem[{{Marchesi} {et~al.}(2019){Marchesi}, {Ajello}, {Zhao}, {Marcotulli},
  {Balokovi{\'c}}, {Brightman}, {Comastri}, {Cusumano}, {Lanzuisi}, {La
  Parola}, {Segreto}, \& {Vignali}}]{marchesi+2019-cftor}
{Marchesi}, S., {Ajello}, M., {Zhao}, X., {et~al.} 2019, \apj, 872, 8

\bibitem[{{Markowitz} {et~al.}(2014){Markowitz}, {Krumpe}, \&
  {Nikutta}}]{markowitz+2014}
{Markowitz}, A.~G., {Krumpe}, M., \& {Nikutta}, R. 2014, \mnras, 439, 1403

\bibitem[{{Mewe} {et~al.}(1995){Mewe}, {Kaastra}, \&
  {Liedahl}}]{mewe+1995-mekal}
{Mewe}, R., {Kaastra}, J.~S., \& {Liedahl}, D.~A. 1995, Legacy, 6, 16

\bibitem[{{Mitsuda} {et~al.}(2007){Mitsuda}, {Bautz}, {Inoue}, {Kelley},
  {Koyama}, {Kunieda}, {Makishima}, {Ogawara}, {Petre}, {Takahashi}, {Tsunemi},
  {White}, {Anabuki}, {Angelini}, {Arnaud}, {Awaki}, {Bamba}, {Boyce}, {Brown},
  {Chan}, {Cottam}, {Dotani}, {Doty}, {Ebisawa}, {Ezoe}, {Fabian}, {Figueroa},
  {Fujimoto}, {Fukazawa}, {Furusho}, {Furuzawa}, {Gendreau}, {Griffiths},
  {Haba}, {Hamaguchi}, {Harrus}, {Hasinger}, {Hatsukade}, {Hayashida}, {Henry},
  {Hiraga}, {Holt}, {Hornschemeier}, {Hughes}, {Hwang}, {Ishida}, {Ishisaki},
  {Isobe}, {Itoh}, {Iyomoto}, {Kahn}, {Kamae}, {Katagiri}, {Kataoka},
  {Katayama}, {Kawai}, {Kilbourne}, {Kinugasa}, {Kissel}, {Kitamoto}, {Kohama},
  {Kohmura}, {Kokubun}, {Kotani}, {Kotoku}, {Kubota}, {Madejski}, {Maeda},
  {Makino}, {Markowitz}, {Matsumoto}, {Matsumoto}, {Matsuoka}, {Matsushita},
  {McCammon}, {Mihara}, {Misaki}, {Miyata}, {Mizuno}, {Mori}, {Mori}, {Morii},
  {Moseley}, {Mukai}, {Murakami}, {Murakami}, {Mushotzky}, {Nagase}, {Namiki},
  {Negoro}, {Nakazawa}, {Nousek}, {Okajima}, {Ogasaka}, {Ohashi}, {Oshima},
  {Ota}, {Ozaki}, {Ozawa}, {Parmar}, {Pence}, {Porter}, {Reeves}, {Ricker},
  {Sakurai}, {Sanders}, {Senda}, {Serlemitsos}, {Shibata}, {Soong}, {Smith},
  {Suzuki}, {Szymkowiak}, {Takahashi}, {Tamagawa}, {Tamura}, {Tamura},
  {Tanaka}, {Tashiro}, {Tawara}, {Terada}, {Terashima}, {Tomida}, {Torii},
  {Tsuboi}, {Tsujimoto}, {Tsuru}, {Turner}, {Ueda}, {Ueno}, {Ueno}, {Uno},
  {Urata}, {Watanabe}, {Yamamoto}, {Yamaoka}, {Yamasaki}, {Yamashita},
  {Yamauchi}, {Yamauchi}, {Yaqoob}, {Yonetoku}, \&
  {Yoshida}}]{mitsuda+2007-suzaku}
{Mitsuda}, K., {Bautz}, M., {Inoue}, H., {et~al.} 2007, \pasj, 59, S1

\bibitem[{{Murphy} \& {Yaqoob}(2009)}]{murphy+yaqoob-2009}
{Murphy}, K.~D., \& {Yaqoob}, T. 2009, \mnras, 397, 1549

\bibitem[{{Nakahara} {et~al.}(2020){Nakahara}, {Doi}, {Murata}, {Nakamura},
  {Hada}, {Asada}, {Sawada-Satoh}, \& {Kameno}}]{nakahara+2020-ngc1052}
{Nakahara}, S., {Doi}, A., {Murata}, Y., {et~al.} 2020, \aj, 159, 14

\bibitem[{{Nandra}(2006)}]{nandra-2006}
{Nandra}, K. 2006, \mnras, 368, L62

\bibitem[{{Netzer}(2015)}]{netzer-2015}
{Netzer}, H. 2015, \araa, 53, 365

\bibitem[{{Ogawa} {et~al.}(2021){Ogawa}, {Ueda}, {Tanimoto}, \&
  {Yamada}}]{ogawa+2021}
{Ogawa}, S., {Ueda}, Y., {Tanimoto}, A., \& {Yamada}, S. 2021, \apj, 906, 84

\bibitem[{{Oh} {et~al.}(2018){Oh}, {Koss}, {Markwardt}, {Schawinski},
  {Baumgartner}, {Barthelmy}, {Cenko}, {Gehrels}, {Mushotzky}, {Petulante},
  {Ricci}, {Lien}, \& {Trakhtenbrot}}]{oh+2018-bat105}
{Oh}, K., {Koss}, M., {Markwardt}, C.~B., {et~al.} 2018, \apjs, 235, 4

\bibitem[{{Osorio-Clavijo} {et~al.}(2020){Osorio-Clavijo},
  {Gonz{\'a}lez-Mart{\'\i}n}, {Papadakis}, {Masegosa}, \&
  {Hern{\'a}ndez-Garc{\'\i}a}}]{osorioClavijo+2020-ngc1052}
{Osorio-Clavijo}, N., {Gonz{\'a}lez-Mart{\'\i}n}, O., {Papadakis}, I.~E.,
  {Masegosa}, J., \& {Hern{\'a}ndez-Garc{\'\i}a}, L. 2020, \mnras, 491, 29

\bibitem[{{Panagiotou} \& {Walter}(2019)}]{panagiotou+walter-2019}
{Panagiotou}, C., \& {Walter}, R. 2019, \aap, 626, A40

\bibitem[{{Parmar} {et~al.}(1997){Parmar}, {Martin}, {Bavdaz}, {Favata},
  {Kuulkers}, {Vacanti}, {Lammers}, {Peacock}, \&
  {Taylor}}]{parmar+1997-bepposaxlecs}
{Parmar}, A.~N., {Martin}, D.~D.~E., {Bavdaz}, M., {et~al.} 1997, \aaps, 122,
  309

\bibitem[{{Ramos Almeida} \& {Ricci}(2017)}]{ramosAlmeida+ricci-2017}
{Ramos Almeida}, C., \& {Ricci}, C. 2017, Nature Astronomy, 1, 679

\bibitem[{{Ricci} {et~al.}(2017{\natexlab{a}}){Ricci}, {Trakhtenbrot}, {Koss},
  {Ueda}, {Del Vecchio}, {Treister}, {Schawinski}, {Paltani}, {Oh}, {Lamperti},
  {Berney}, {Gandhi}, {Ichikawa}, {Bauer}, {Ho}, {Asmus}, {Beckmann}, {Soldi},
  {Balokovi{\'c}}, {Gehrels}, \& {Markwardt}}]{ricci+2017-bass}
{Ricci}, C., {Trakhtenbrot}, B., {Koss}, M.~J., {et~al.} 2017{\natexlab{a}},
  \apjs, 233, 17

\bibitem[{{Ricci} {et~al.}(2017{\natexlab{b}}){Ricci}, {Trakhtenbrot}, {Koss},
  {Ueda}, {Schawinski}, {Oh}, {Lamperti}, {Mushotzky}, {Treister}, {Ho},
  {Weigel}, {Bauer}, {Paltani}, {Fabian}, {Xie}, \&
  {Gehrels}}]{ricci+2017-cftor}
---. 2017{\natexlab{b}}, \nat, 549, 488

\bibitem[{{Risaliti} {et~al.}(2002){Risaliti}, {Elvis}, \&
  {Nicastro}}]{risaliti+2002}
{Risaliti}, G., {Elvis}, M., \& {Nicastro}, F. 2002, \apj, 571, 234

\bibitem[{{Rivers} {et~al.}(2013){Rivers}, {Markowitz}, \&
  {Rothschild}}]{rivers+2013}
{Rivers}, E., {Markowitz}, A., \& {Rothschild}, R. 2013, \apj, 772, 114

\bibitem[{{Sambruna} {et~al.}(2006){Sambruna}, {Gliozzi}, {Tavecchio},
  {Maraschi}, \& {Foschini}}]{sambruna+2006}
{Sambruna}, R.~M., {Gliozzi}, M., {Tavecchio}, F., {Maraschi}, L., \&
  {Foschini}, L. 2006, \apj, 652, 146

\bibitem[{{Sawada-Satoh} {et~al.}(2008){Sawada-Satoh}, {Kameno}, {Nakamura},
  {Namikawa}, {Shibata}, \& {Inoue}}]{sawadaSatoh+2008-ngc1052}
{Sawada-Satoh}, S., {Kameno}, S., {Nakamura}, K., {et~al.} 2008, \apj, 680, 191

\bibitem[{{Sawada-Satoh} {et~al.}(2016){Sawada-Satoh}, {Roh}, {Oh}, {Lee},
  {Byun}, {Kameno}, {Yeom}, {Jung}, {Kim}, \&
  {Hwang}}]{sawadaSatoh+2016-ngc1052}
{Sawada-Satoh}, S., {Roh}, D.-G., {Oh}, S.-J., {et~al.} 2016, \apjl, 830, L3

\bibitem[{{Sawada-Satoh} {et~al.}(2019){Sawada-Satoh}, {Byun}, {Lee}, {Oh},
  {Roh}, {Kameno}, {Yeom}, {Jung}, {Oh}, \& {Kim}}]{sawadaSatoh+2019-ngc1052}
{Sawada-Satoh}, S., {Byun}, D.-Y., {Lee}, S.-S., {et~al.} 2019, \apjl, 872, L21

\bibitem[{{She} {et~al.}(2018){She}, {Ho}, {Feng}, \& {Cui}}]{she+2018}
{She}, R., {Ho}, L.~C., {Feng}, H., \& {Cui}, C. 2018, \apj, 859, 152

\bibitem[{{Shu} {et~al.}(2011){Shu}, {Yaqoob}, \& {Wang}}]{shu+2011}
{Shu}, X.~W., {Yaqoob}, T., \& {Wang}, J.~X. 2011, \apj, 738, 147

\bibitem[{{Stalevski} {et~al.}(2016){Stalevski}, {Ricci}, {Ueda}, {Lira},
  {Fritz}, \& {Baes}}]{stalevski+2016}
{Stalevski}, M., {Ricci}, C., {Ueda}, Y., {et~al.} 2016, \mnras, 458, 2288

\bibitem[{{Str{\"u}der} {et~al.}(2001){Str{\"u}der}, {Briel}, {Dennerl},
  {Hartmann}, {Kendziorra}, {Meidinger}, {Pfeffermann}, {Reppin}, {Aschenbach},
  {Bornemann}, {Br{\"a}uninger}, {Burkert}, {Elender}, {Freyberg}, {Haberl},
  {Hartner}, {Heuschmann}, {Hippmann}, {Kastelic}, {Kemmer}, {Kettenring},
  {Kink}, {Krause}, {M{\"u}ller}, {Oppitz}, {Pietsch}, {Popp}, {Predehl},
  {Read}, {Stephan}, {St{\"o}tter}, {Tr{\"u}mper}, {Holl}, {Kemmer}, {Soltau},
  {St{\"o}tter}, {Weber}, {Weichert}, {von Zanthier}, {Carathanassis}, {Lutz},
  {Richter}, {Solc}, {B{\"o}ttcher}, {Kuster}, {Staubert}, {Abbey}, {Holland},
  {Turner}, {Balasini}, {Bignami}, {La Palombara}, {Villa}, {Buttler},
  {Gianini}, {Lain{\'e}}, {Lumb}, \& {Dhez}}]{strueder+2001-epicpn}
{Str{\"u}der}, L., {Briel}, U., {Dennerl}, K., {et~al.} 2001, \aap, 365, L18

\bibitem[{{Takahashi} {et~al.}(2007){Takahashi}, {Abe}, {Endo}, {Endo}, {Ezoe},
  {Fukazawa}, {Hamaya}, {Hirakuri}, {Hong}, {Horii}, {Inoue}, {Isobe}, {Itoh},
  {Iyomoto}, {Kamae}, {Kasama}, {Kataoka}, {Kato}, {Kawaharada}, {Kawano},
  {Kawashima}, {Kawasoe}, {Kishishita}, {Kitaguchi}, {Kobayashi}, {Kokubun},
  {Kotoku}, {Kouda}, {Kubota}, {Kuroda}, {Madejski}, {Makishima}, {Masukawa},
  {Matsumoto}, {Mitani}, {Miyawaki}, {Mizuno}, {Mori}, {Mori}, {Murashima},
  {Murakami}, {Nakazawa}, {Niko}, {Nomachi}, {Okada}, {Ohno}, {Oonuki}, {Ota},
  {Ozawa}, {Sato}, {Shinoda}, {Sugiho}, {Suzuki}, {Taguchi}, {Takahashi},
  {Takahashi}, {Takeda}, {Tamura}, {Tamura}, {Tanaka}, {Tanihata}, {Tashiro},
  {Terada}, {Tominaga}, {Uchiyama}, {Watanabe}, {Yamaoka}, {Yanagida}, \&
  {Yonetoku}}]{takahashi+2007-suzakuhxd}
{Takahashi}, T., {Abe}, K., {Endo}, M., {et~al.} 2007, \pasj, 59, 35

\bibitem[{{Tueller} {et~al.}(2010){Tueller}, {Baumgartner}, {Markwardt},
  {Skinner}, {Mushotzky}, {Ajello}, {Barthelmy}, {Beardmore}, {Brandt},
  {Burrows}, {Chincarini}, {Campana}, {Cummings}, {Cusumano}, {Evans},
  {Fenimore}, {Gehrels}, {Godet}, {Grupe}, {Holland}, {Kennea}, {Krimm},
  {Koss}, {Moretti}, {Mukai}, {Osborne}, {Okajima}, {Pagani}, {Page}, {Palmer},
  {Parsons}, {Schneider}, {Sakamoto}, {Sambruna}, {Sato}, {Stamatikos},
  {Stroh}, {Ukwata}, \& {Winter}}]{tueller+2010-bat22}
{Tueller}, J., {Baumgartner}, W.~H., {Markwardt}, C.~B., {et~al.} 2010, \apjs,
  186, 378

\bibitem[{{Turner} {et~al.}(2001){Turner}, {Abbey}, {Arnaud}, {Balasini},
  {Barbera}, {Belsole}, {Bennie}, {Bernard}, {Bignami}, {Boer}, {Briel},
  {Butler}, {Cara}, {Chabaud}, {Cole}, {Collura}, {Conte}, {Cros}, {Denby},
  {Dhez}, {Di Coco}, {Dowson}, {Ferrando}, {Ghizzardi}, {Gianotti}, {Goodall},
  {Gretton}, {Griffiths}, {Hainaut}, {Hochedez}, {Holland}, {Jourdain},
  {Kendziorra}, {Lagostina}, {Laine}, {La Palombara}, {Lortholary}, {Lumb},
  {Marty}, {Molendi}, {Pigot}, {Poindron}, {Pounds}, {Reeves}, {Reppin},
  {Rothenflug}, {Salvetat}, {Sauvageot}, {Schmitt}, {Sembay}, {Short},
  {Spragg}, {Stephen}, {Str{\"u}der}, {Tiengo}, {Trifoglio}, {Tr{\"u}mper},
  {Vercellone}, {Vigroux}, {Villa}, {Ward}, {Whitehead}, \&
  {Zonca}}]{turner+2001-epicmos}
{Turner}, M.~J.~L., {Abbey}, A., {Arnaud}, M., {et~al.} 2001, \aap, 365, L27

\bibitem[{{Ubertini} {et~al.}(2003){Ubertini}, {Lebrun}, {Di Cocco}, {Bazzano},
  {Bird}, {Broenstad}, {Goldwurm}, {La Rosa}, {Labanti}, {Laurent}, {Mirabel},
  {Quadrini}, {Ramsey}, {Reglero}, {Sabau}, {Sacco}, {Staubert}, {Vigroux},
  {Weisskopf}, \& {Zdziarski}}]{ubertini+2003-ibis}
{Ubertini}, P., {Lebrun}, F., {Di Cocco}, G., {et~al.} 2003, \aap, 411, L131

\bibitem[{{Urry} \& {Padovani}(1995)}]{urry+padovani-1995}
{Urry}, C.~M., \& {Padovani}, P. 1995, \pasp, 107, 803

\bibitem[{{Ursini} {et~al.}(2015){Ursini}, {Marinucci}, {Matt}, {Bianchi},
  {Tortosa}, {Stern}, {Ar{\'e}valo}, {Ballantyne}, {Bauer}, {Fabian},
  {Harrison}, {Lohfink}, {Reynolds}, \& {Walton}}]{ursini+2015-ngc7213}
{Ursini}, F., {Marinucci}, A., {Matt}, G., {et~al.} 2015, \mnras, 452, 3266

\bibitem[{{Uttley} {et~al.}(2019){Uttley}, {den Hartog}, {Bambi}, {Barret},
  {Bianchi}, {Bursa}, {Cappi}, {Casella}, {Cash}, {Costantini}, {Dauser}, {Diaz
  Trigo}, {Gendreau}, {Grinberg}, {den Herder}, {Ingram}, {Kara}, {Markoff},
  {Mingo}, {Panessa}, {Poppenh{\"a}ger}, {R{\'o}{\.z}a{\'n}ska}, {Svoboda},
  {Wijers}, {Willingale}, {Wilms}, \& {Wise}}]{uttley+2019-wp}
{Uttley}, P., {den Hartog}, R., {Bambi}, C., {et~al.} 2019, arXiv e-prints,
  arXiv:1908.03144

\bibitem[{{Vasudevan} {et~al.}(2013){Vasudevan}, {Brandt}, {Mushotzky},
  {Winter}, {Baumgartner}, {Shimizu}, {Schneider}, \&
  {Nousek}}]{vasudevan+2013}
{Vasudevan}, R.~V., {Brandt}, W.~N., {Mushotzky}, R.~F., {et~al.} 2013, \apj,
  763, 111

\bibitem[{{Vermeulen} {et~al.}(2003){Vermeulen}, {Ros}, {Kellermann}, {Cohen},
  {Zensus}, \& {van Langevelde}}]{vermeulen+2003-ngc1052}
{Vermeulen}, R.~C., {Ros}, E., {Kellermann}, K.~I., {et~al.} 2003, \aap, 401,
  113

\bibitem[{{Wajima} {et~al.}(2020){Wajima}, {Kino}, \&
  {Kawakatu}}]{wajima+2020-ngc1275}
{Wajima}, K., {Kino}, M., \& {Kawakatu}, N. 2020, \apj, 895, 35

\bibitem[{{Walter} {et~al.}(2010){Walter}, {Rohlfs}, {Meharga}, {Binko},
  {Morisset}, {Beck}, {Produit}, {Pavan}, {Savchenko}, {Ferrigno},
  {Frankowski}, \& {Bordas}}]{walter+2010-isdc}
{Walter}, R., {Rohlfs}, R., {Meharga}, M.~T., {et~al.} 2010, in Eighth Integral
  Workshop. The Restless Gamma-ray Universe (INTEGRAL 2010), 162

\bibitem[{{Weaver} {et~al.}(1999){Weaver}, {Wilson}, {Henkel}, \&
  {Braatz}}]{weaver+1999-ngc1052}
{Weaver}, K.~A., {Wilson}, A.~S., {Henkel}, C., \& {Braatz}, J.~A. 1999, The
  Astrophysical Journal, 520, 130

\bibitem[{{Winkler} {et~al.}(2003){Winkler}, {Courvoisier}, {Di Cocco},
  {Gehrels}, {Gim{\'e}nez}, {Grebenev}, {Hermsen}, {Mas-Hesse}, {Lebrun},
  {Lund}, {Palumbo}, {Paul}, {Roques}, {Schnopper}, {Sch{\"o}nfelder},
  {Sunyaev}, {Teegarden}, {Ubertini}, {Vedrenne}, \&
  {Dean}}]{winkler+2003-integral}
{Winkler}, C., {Courvoisier}, T.~J.-L., {Di Cocco}, G., {et~al.} 2003, \aap,
  411, L1

\bibitem[{{Yaqoob} {et~al.}(2015){Yaqoob}, {Tatum}, {Scholtes}, {Gottlieb}, \&
  {Turner}}]{yaqoob+2015-mrk3}
{Yaqoob}, T., {Tatum}, M.~M., {Scholtes}, A., {Gottlieb}, A., \& {Turner},
  T.~J. 2015, \mnras, 454, 973

\bibitem[{{Younes} {et~al.}(2019){Younes}, {Ptak}, {Ho}, {Xie}, {Terasima},
  {Yuan}, {Huppenkothen}, \& {Yukita}}]{younes+2019}
{Younes}, G., {Ptak}, A., {Ho}, L.~C., {et~al.} 2019, \apj, 870, 73

\bibitem[{{Young} {et~al.}(2018){Young}, {McHardy}, {Emmanoulopoulos}, \&
  {Connolly}}]{young+2018-m81}
{Young}, A.~J., {McHardy}, I., {Emmanoulopoulos}, D., \& {Connolly}, S. 2018,
  \mnras, 476, 5698

\bibitem[{{Zaino} {et~al.}(2020){Zaino}, {Bianchi}, {Marinucci}, {Matt},
  {Bauer}, {Brandt}, {Gandhi}, {Guainazzi}, {Iwasawa}, {Puccetti}, {Ricci}, \&
  {Walton}}]{zaino+2020-ngc1068}
{Zaino}, A., {Bianchi}, S., {Marinucci}, A., {et~al.} 2020, \mnras, 492, 3872

\bibitem[{{Zdziarski} {et~al.}(1996){Zdziarski}, {Johnson}, \&
  {Magdziarz}}]{zdziarski+1996}
{Zdziarski}, A.~A., {Johnson}, W.~N., \& {Magdziarz}, P. 1996, \mnras, 283, 193

\bibitem[{{Zhao} {et~al.}(2020){Zhao}, {Marchesi}, {Ajello}, {Balokovi{\'c}},
  \& {Fischer}}]{zhao+2020}
{Zhao}, X., {Marchesi}, S., {Ajello}, M., {Balokovi{\'c}}, M., \& {Fischer}, T.
  2020, \apj, 894, 71

\bibitem[{{Zhao} {et~al.}(2021){Zhao}, {Marchesi}, {Ajello}, {Cole}, {Hu},
  {Silver}, \& {Torres-Alb{\`a}}}]{zhao+2021}
{Zhao}, X., {Marchesi}, S., {Ajello}, M., {et~al.} 2021, \aap, 650, A57

\bibitem[{{{\.Z}ycki} {et~al.}(1999){{\.Z}ycki}, {Done}, \&
  {Smith}}]{zycki+1999}
{{\.Z}ycki}, P.~T., {Done}, C., \& {Smith}, D.~A. 1999, \mnras, 309, 561

\end{thebibliography}

\end{document}